\documentclass{aa}  

\usepackage{graphicx}
\usepackage{txfonts}

\usepackage{subcaption}  
\usepackage{lscape}         
\usepackage{placeins}    
\usepackage{makecell}
\usepackage{orcidlink}
\usepackage{CJKutf8}

\newcommand{\chinesename}[1]{\begin{CJK*}{UTF8}{gbsn}#1\end{CJK*}}

\begin{document}
\hypersetup{hidelinks}

   \title{Prospects for detecting hot-Jupiter tidal decay\\through transit timing with the \textit{Earth 2.0} mission}

   \author{Wenqin Wang (\chinesename{汪文沁})\orcidlink{0009-0007-2398-9076}\inst{1,2}
          \and
          Yue Wang (\chinesename{王越})\orcidlink{0009-0002-2422-3930}\inst{1,2}
          \and
          Di-Chang Chen (\chinesename{陈迪昌})\orcidlink{0000-0003-0707-3213}\inst{1,2}
          \thanks{Corresponding author: \protect\email{chendch28@mail.sysu.edu.cn}}
          \and
          Hui Zhang (\chinesename{张辉})\inst{3}
          \and
          Jian~Ge (\chinesename{葛健})\inst{3}
          \and
          Bo Ma (\chinesename{马波})\orcidlink{0000-0002-0378-2023}\inst{1,2}
          \thanks{Corresponding author: \protect\email{mabo8@mail.sysu.edu.cn}}
          }

   \institute{School of Physics and Astronomy,
              Sun Yat-sen University,
              Zhuhai 519082, China 
         \and
             Chinese Space Station Survey Telescope (CSST) Science Center for the Guangdong-Hongkong-Macau Greater Bay Area,
             Sun Yat-sen University,
             Zhuhai 519082, China
        \and
             Shanghai Astronomical Observatory,
             Chinese Academy of Sciences,
             Shanghai 200030, China
             }

   \date{Received 24 July 2026 / Accepted 30 October 2026}

   \titlerunning{Detecting hot-Jupiter tidal decay with ET}

    \authorrunning{Wang et al.}

  \abstract
{Orbital decay of hot Jupiters provides a unique opportunity to probe tidal dissipation in star-planet systems. However, despite extensive observational efforts, only one hot Jupiter, namely WASP-12b, currently exhibits conclusive observational evidence for orbital decay. 
Expanding this sample is therefore crucial for establishing a statistical understanding of tidal interactions in planetary systems. }
{Long-term transit timing variations (TTVs) provide a direct observational probe of orbital decay induced by tidal dissipation between a hot Jupiter and its host star. The \textit{Earth 2.0} (ET) mission will revisit the original \textit{Kepler} field, delivering high-precision transit timing measurements and extending the observational baseline to approximately 24 years (2009--2033).
In this work, we aim to assess the capability of ET to detect orbital decay in hot Jupiters previously observed by \textit{Kepler}.}
{We generate four years of simulated ET light curves based on the expected photometric precision and observing cadence of the mission and analyze them jointly with archival \textit{Kepler} and TESS observations.
Using these combined datasets, we perform long-term TTV analyses and predict the achievable precision for measuring orbital period variations.} 
{We find that ET can substantially improve the precision of orbital decay measurements in hot Jupiter systems by a factor of 7.2. 

Assuming the stellar tidal quality factor inferred from previous observations, ET is expected to detect orbital decay in three additional hot Jupiter systems.
}
{

The expanded sample of hot Jupiters with measured orbital decay will establish ET as a powerful probe of tidal dissipation in star–planet systems, providing tighter constraints on stellar tidal quality factors and new insights into the tidal evolution of hot Jupiters.
}


   \maketitle
\nolinenumbers

\section{Introduction}

Hot Jupiters are gas giant planets on short-period orbits, which provide a unique laboratory for studying tidal interactions between stars and planets \citep{Hansen2010, Dawson2018}.
At such close separations, tidal dissipation within the host star is expected to drive long-term orbital evolution, potentially leading to measurable orbital decay \citep{Rasio1996, Levrard2009}. 
However, the efficiency of this dissipation process remains poorly constrained from direct observations \citep{Ogilvie2014, Weinberg2017, Yee2020}.

The most direct observable signature of tidal decay is a secular change in the orbital period, $\dot{P} = \mathrm{d}P/\mathrm{d}t$, which is directly linked to the stellar tidal dissipation efficiency \citep{Ogilvie2014}.

This efficiency is commonly parameterized by the modified tidal quality factor $Q'_{\star}$, which provides an effective measure of the overall tidal energy dissipation in the star.
Several physical mechanisms can contribute to this dissipation, including turbulent dissipation of equilibrium tides in convective regions and the excitation and damping of dynamical tides, such as inertial waves in convective zones and internal gravity waves in radiative regions \citep{Zahn1977, Zahn1989, Ogilvie2014, Penev2018}.
The efficiency of these mechanisms can vary by several orders of magnitude and depends strongly on stellar structure, rotation, and the tidal forcing frequency; therefore, $Q'_{\star}$ should be regarded as an effective parameter describing their combined action \citep{Ogilvie2014}.
This efficiency determines the inspiral timescale of close-in planets \citep{Maxinyue2025}.
Measuring $\dot{P}$ across multiple systems therefore provides a key empirical pathway to constraining tidal dissipation theory \citep{Hansen2010, Maciejewski2016} and the long-term evolution of hot Jupiters.

Despite its importance, robust detections of tidal orbital decay remain extremely rare. 
Transit timing variations (TTVs) offer the primary method for directly measuring $\dot{P}$ and have been widely applied to study hot Jupiters and warm-Neptunes in our ExoEcho project (\citealp{Wang2024}; \citealp{Zhangzixin24}, \citeyear{Zhangzixin26}; \citealp{Wu2026}).
However, long-term TTV measurements face significant challenges. 
Existing datasets are heterogeneous in both precision and temporal coverage \citep{Wang2024}, and apparent timing drifts can be mimicked by other effects, particularly the light-travel time effect (LTTE) induced by additional companions \citep{Bouma2020}.
Indeed, several previously reported decay signals have later been reinterpreted as LTTE, such as WASP-4~b and Kepler-1658~b \citep{Winn2025}.
To date, WASP-12~b remains the only system with a secure detection of orbital decay, with $\dot{P} \approx -29 \pm 2~\mathrm{ms\,yr^{-1}}$ \citep{Maciejewski2016, Patra2017, Yee2020, Maxinyue2025}. 
Interpreting this signal as tidal dissipation implies a relatively low stellar tidal quality factor, $Q'_\star \sim 10^5$ \citep{Turner2021}.
Whether this value is representative of the broader hot Jupiter population or instead an outlier remains unclear.

Detecting tidal decay in additional systems requires both high timing precision and long observational baselines \citep{Maxinyue2025, Wang2025}. 
While the \textit{Kepler} mission provides excellent single transit timing precision, its $\sim$4-year duration was insufficient \citep{Borucki2010}.
TESS extends the time coverage but with lower photometric precision \citep{Ricker2015}. 
For typical decay rates ($|\dot{P}| \lesssim 10~\mathrm{ms\,yr^{-1}}$), a temporal baseline exceeding $\sim$15 years is generally required to robustly distinguish tidal decay from LTTE-induced signals \citep{Patra2020, Bouma2020, Wang2025}.

The \textit{Earth 2.0} (ET) Space Mission is a next-generation space-based transit survey \citep{Ge2022a, Ge2022b, Ge2024}. Equipped with six transit telescopes with 28~cm aperture size, it is designed to achieve \textit{Kepler}-like photometric precision while re-observing the original \textit{Kepler} field over a new four-year baseline.
By combining ET observations with archival \textit{Kepler} and TESS data, the total timing baseline for many systems will exceed 20~years, opening a promising window for detecting weak orbital decay signals.
However, the sensitivity of such combined datasets to $\dot{P}$ has not yet been quantified. 
In particular, it remains unclear how observational factors such as photometric precision, cadence, and temporal sampling affect the detectability of small quadratic timing trends, and to what extent ET can break degeneracies with LTTE signals. 
A systematic simulation study is therefore required to assess the detectability of tidal decay under realistic observational conditions and to evaluate the scientific return of the ET mission in this context.

In this work, we perform a comprehensive simulation study to evaluate the improvement in $\dot{P}$ constraints enabled by the ET mission. 
Building on our previous work on exomoon detectability with ET \citep{Cuizimiao2026}, we extend the simulation framework to tidal orbital decay. 
We construct realistic transit timing datasets by combining archival \textit{Kepler} and TESS observations with simulated ET data, and assess the resulting precision in measuring $\dot{P}$.
Our goal is to provide quantitative predictions for the number of detectable systems and the achievable constraints on the stellar tidal quality factor $Q'_\star$, thereby informing both future observational strategies and theoretical models.

This paper is organized as follows. Section~\ref{sec:Data} describes the target selection and transit timing data processing and simulation; Section~\ref{sec:Timing} presents the secular TTV model and analysis method; Section~\ref{sec:Results} reports the TTV fitting results and predictions of newly identified hot Jupiters with orbital decay, and discusses the implications. 
Section~\ref{sec:Conclusion} summarizes our main conclusions. 

\section{Target selection and transit timing data} \label{sec:Data}
\subsection{Target selection}\label{sec:Target}

We selected 28 hot Jupiters (planet masses of 0.35–13~$M_{\rm Jup}$, and orbital periods $P < 10$~days) originally discovered by the \textit{Kepler} mission.
This sample comprises well-characterized short-period giant planets with high-quality archival transit data, enabling robust timing analysis in the \textit{Kepler} and \textit{ET} field.
The targets and their basic stellar and planetary parameters from the literature are listed in Table~\ref{tab:sample}.

Kepler-91~b was excluded from our sample because its host star is a red giant with a large stellar radius, producing a relatively shallow transit signal that is strongly affected by stellar variability and correlated (red) noise \citep{Lillo-Box2014L}. Although detrending can remove long-term trends, the residual noise remains comparable to the transit signal, resulting in large uncertainties in the measured transit mid-times.

\subsection{\textit{Kepler} and TESS archival data} \label{sec:Kepler}

\begin{figure*}[ht!]
    \centering
    \includegraphics[width=0.95\textwidth]{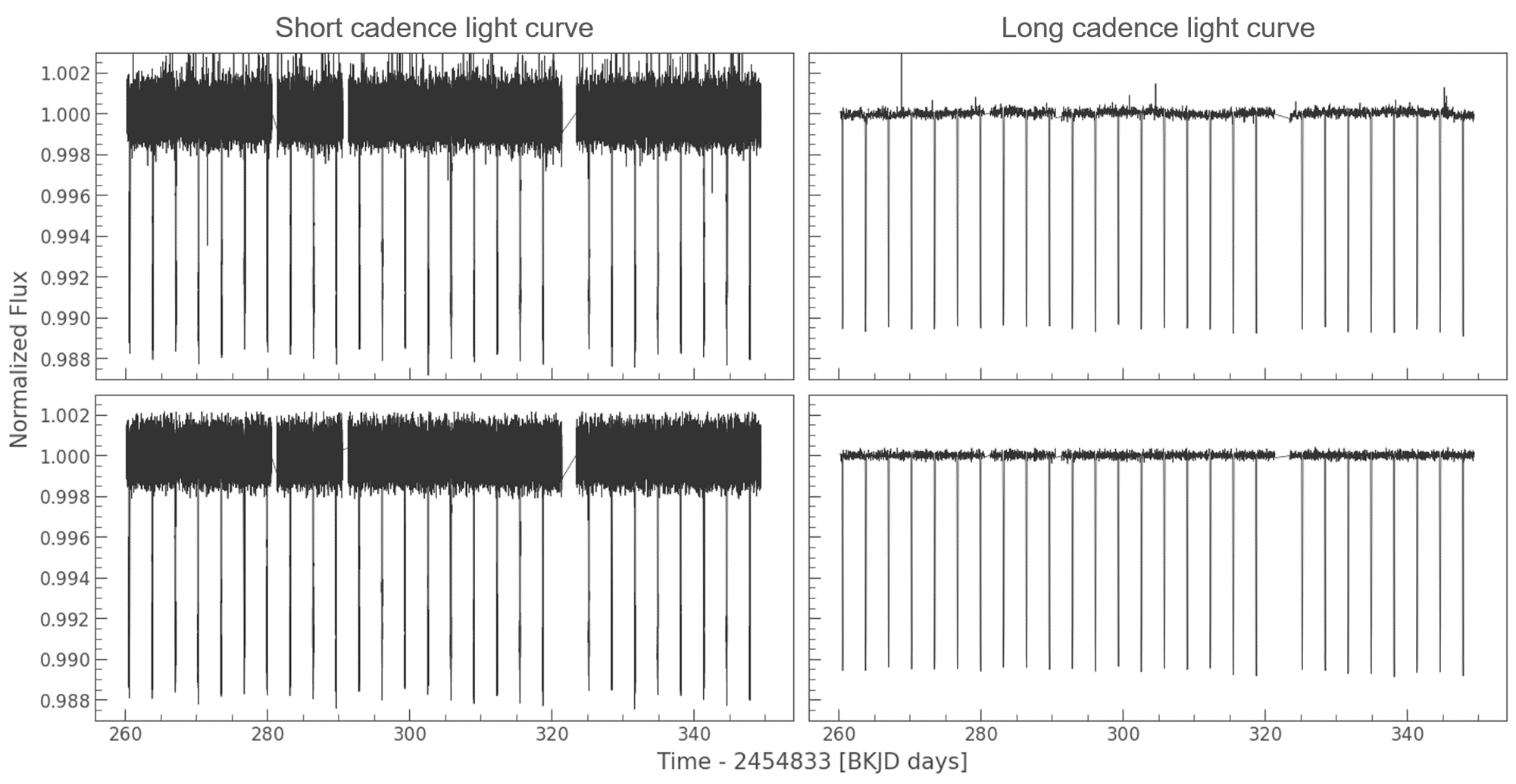}
    \caption{
    Comparison of the Kepler-6 b light curves before and after detrending. 
    The left panels show the short-cadence (SC) data from quarter~3, with the raw light curve (before detrending) shown at the top and the detrended version at the bottom. 
    The right panels show the corresponding long-cadence (LC) data from the same quarter. Outliers (spikes) were removed using a sigma-clipping procedure.
    The figure illustrates the improvement in baseline stability and transit shape clarity after applying the detrending process.
    }
    \label{fig:kepler_detrend}
\end{figure*}

We downloaded \textit{Kepler} light curves from the Mikulski Archive for Space Telescopes (MAST) using the \texttt{lightkurve} package \citep{2018ascl.soft12013L}. 
For each target, both short-cadence (SC, 1~min) and long-cadence (LC, 30~min) data were retrieved.
We also obtained 2-minute cadence \textit{TESS} light curves from MAST.
The raw light curves were detrended using \texttt{lightkurve} to remove long-term systematics and stellar variability. An example of the detrending procedure is shown in Fig.~\ref{fig:kepler_detrend}.

\begin{figure}[ht!]
    \centering
    \includegraphics[width=0.48\textwidth]{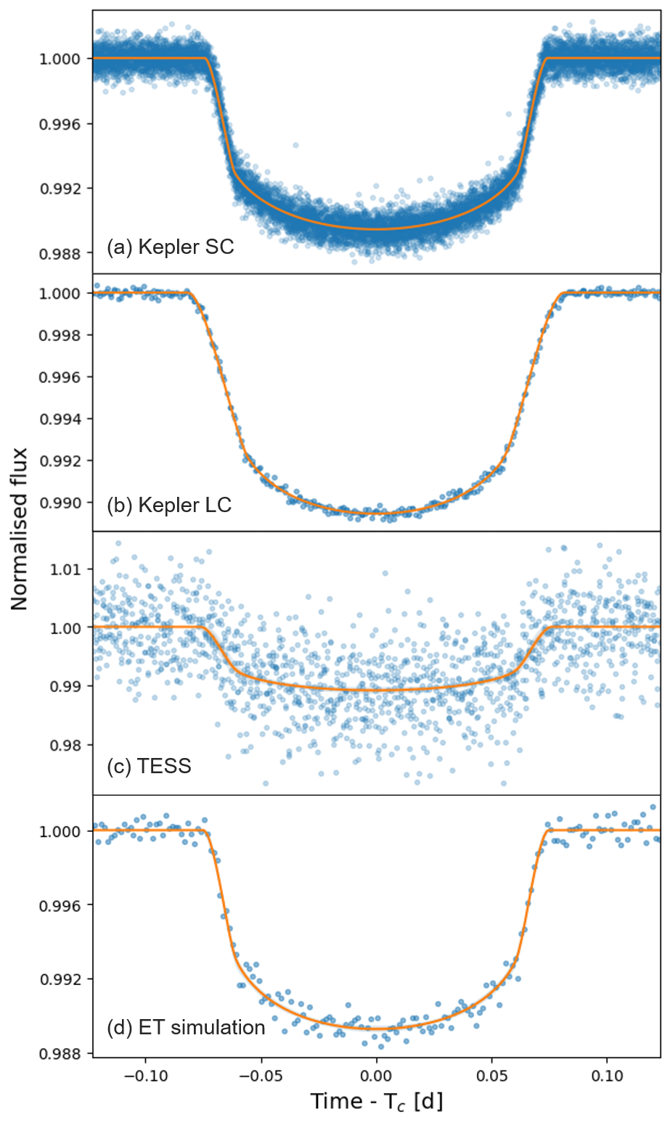}
    \caption{
    Phase-folded transit light curves of Kepler-6 b observed by \textit{Kepler} and \textit{TESS}, and a simulated ET transit.
    The panels show the \textit{Kepler} quarter~3 LC data (a), quarter~3 SC data (b), \textit{TESS} sector~41 data (c), and the simulated 2-min cadence ET transit (d).
    The blue points in panels (a)--(c) represent the observed flux, and the orange lines show the best-fit transit models.
    The transit in panel (d) is a simulated single transit generated using the system parameters of Kepler-6 b.
    }
    \label{fig:Kepler_6b_transits}
\end{figure}

The transit modeling was performed using the \texttt{PyTransit} package \citep{Parviainen2015pytransit}, adopting the analytic model of \citet{Mandel2002}. 

The reference epoch $T_0$ and orbital period $P$ were taken from the literature and used to calculate the expected transit times for each epoch (Table~\ref{tab:ephemerides}). Each transit was then fitted individually, with the mid-transit time $T_c$ treated as a free parameter with a normal prior centered on the corresponding expected transit time, while $P$ was assigned a normal prior centered on its literature value.
Uniform priors were adopted for the geometric parameters, including the scaled semi-major axis $a/R_\star$, impact parameter $b$, stellar density $\rho_\star$, and planet-to-star radius ratio $R_p/R_\star$. The orbital eccentricity and argument of periastron were parameterized as $\sqrt{e}\cos\omega$ and $\sqrt{e}\sin\omega$. A quadratic limb-darkening law was adopted, with both coefficients treated as free parameters.
An example of the phase-folded light curves and best-fit models is shown in Fig.~\ref{fig:Kepler_6b_transits}.

For three grazing systems (Kepler-76~b, Kepler-424~b, and Kepler-447~b), the limb-darkening and geometric parameters are highly degenerate. For these systems, we adopted literature values and applied normal priors.
For Kepler-424~b and Kepler-447~b, where limb-darkening coefficients are not available in the literature, we estimated them using the Limb Darkening Toolkit (\texttt{LDTk}; \citealt{Parviainen2015LDTK}) based on the stellar parameters ($T_{\mathrm{eff}}$, $\log g$, [Fe/H]).
\texttt{LDTk} computes model intensities over the \textit{Kepler} bandpass and fits a quadratic law to derive the coefficients $u_1$ and $u_2$.

To obtain robust posterior distributions, we first performed a global optimization using a differential evolution algorithm to identify suitable initial parameter values. This was followed by a Markov Chain Monte Carlo (MCMC) analysis with 100 walkers and 2000 steps per chain.

The best-fit mid-transit times $T_c$ were determined from the medians of the corresponding posterior distributions, with the uncertainties estimated as the standard deviations of the posterior distributions. These values were then used in the subsequent TTV analysis.

During the transit fitting process, we applied additional quality filters to ensure reliable mid-transit time measurements. 
First, we excluded transits that were not fully covered in the light curves (i.e., lacking complete ingress or egress). Second, we removed low-quality transits with weak or noisy signals by retaining only transits with the signal-to-noise ratio (S/N) $>$ 5.
The S/N was defined as
\begin{equation}
\mathrm{S/N} = \frac{|\mathrm{\delta}|}{\sigma_1} \sqrt{N_{\mathrm{in}}}
\end{equation}
where $\delta$ is the transit depth, $\sigma_1$ is the standard deviation of the out-of-transit residuals, and $N_{in}$ is the number of in-transit data points. 
This threshold ensures reliable timing measurements while retaining a sufficient sample size. 

The adopted reference ephemerides, including the transit duration $\Delta_{14}$, and the numbers of usable transits obtained from each instrument are summarized in Table~\ref{tab:ephemerides}.

The measured mid-transit times for all 28 systems are presented in Table~\ref{tab:midtc}, and the full dataset is made available online in machine-readable form.

\subsection{ET data simulation}\label{subsec:ET_Transit}

The ET mission is a next-generation space-based transit survey designed to achieve ultra-high photometric precision in a large field of view \citep{Ge2024}. 
The transit module consists of six 28-cm telescopes operating at the Sun–Earth L2 point. 
During its planned four-year primary mission (2029–2033), ET will continuously monitor more than two million FGKM dwarfs ($G < 16~\mathrm{mag}$), including the original \textit{Kepler} field. 

For each target in our sample, we simulated four years of continuous ET observations using \texttt{PyTransit}. 
Synthetic light curves were generated at a cadence of 2 minutes, adopting the system parameters derived from the \textit{Kepler} SC fits (including $T_0$, $P$, $b$, $\rho_\star$, $R_p/R_\star$, and limb-darkening coefficients), together with the literature-based transit duration $\Delta_{14}$.
$\dot{P}$ was set to the theoretical value predicted for each system assuming $\log Q^{'}_\star = 5.7^{+0.4}_{-0.3}$, as inferred from the temporal evolution of hot Jupiter occurrence rates \citep{Chen2025}. Following the equilibrium tidal formalism of \citet{Goldreich1966}, $\dot{P}$ was computed as
\begin{equation}\label{eq:Pdot}
\dot{P}
=
-\frac{27\pi}{2Q'_\star}
\left(\frac{M_p}{M_\star}\right)
\left(\frac{a}{R_\star}\right)^{-5}.
\end{equation}
The resulting $\dot{P}$ values were incorporated into the simulated transit ephemerides. The stellar and planetary parameters adopted in the calculation are listed in Table~\ref{tab:sample}, and the corresponding theoretical $\dot{P}$ values are given in Table~\ref{tab:timing}.

\begin{figure}[htbp]
    \centering
    \includegraphics[width=0.42\textwidth]{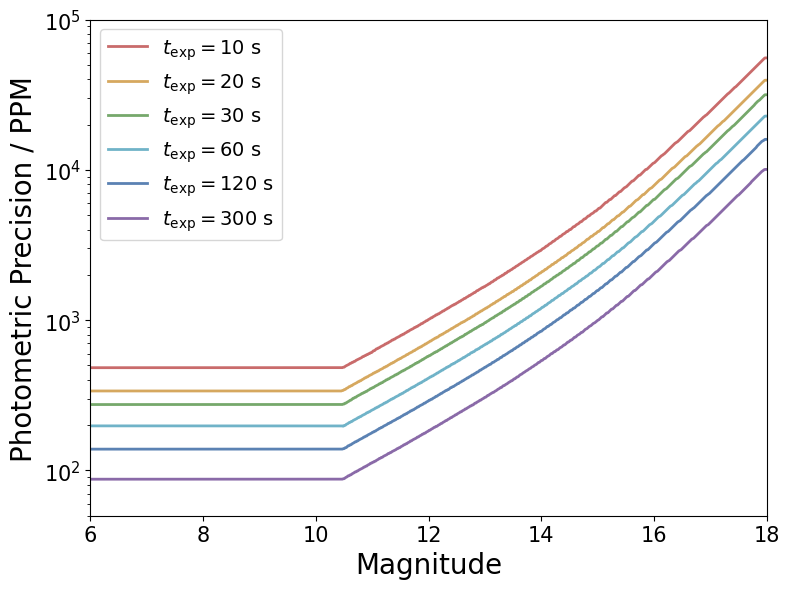}
    \caption{The expected photometric precision of the ET transit module (comprising six 28 cm wide-field telescopes) as a function of target magnitude.
}
    \label{fig:et-precision}
\end{figure}

Instrumental and astrophysical noise were incorporated into the simulated ET light curves using the ET photometric precision model provided by the instrument team (Fig.~\ref{fig:et-precision}).
The \textit{Kepler} magnitude ($K_{\rm p}$), obtained from the NASA Exoplanet Archive, was used to determine the corresponding noise level for each target.
The adopted parameters, $K_{\rm p}$, and number of simulated ET transits are summarized for each system in Table~\ref{tab:transitparams}.

Each system yields on the order of hundreds of transit epochs over the four-year ET mission. The simulated ET light curves were re-analyzed using the same fitting procedure as applied to the real \textit{Kepler} and \textit{TESS} data to extract individual mid-transit times and uncertainties. 
Finally, the ET-derived transit times were combined with our \textit{Kepler} and TESS measurements, extending the total timing baseline to more than two decades and enabling sensitive tests of long-term orbital variations. An example of a simulated ET transit is shown in Fig.~\ref{fig:Kepler_6b_transits} for Kepler-6~b.

\section{Analysis} \label{sec:Timing}

To investigate the presence of orbital period variations, we modeled the combined transit timing datasets from \textit{Kepler} and \textit{TESS}. For the \textit{Kepler} data, both SC and LC light curves are available for most targets. When the same transit epoch is covered by both cadences, we adopt the SC measurement due to its higher precision, ensuring that each epoch contributes only a single timing measurement.

We consider two competing models. The first is a constant-period (linear ephemeris) model,
\begin{equation}
T_N = T_0 +NP,
\end{equation}
where $T_N$ is the mid-transit time of the $N$-th epoch, $T_0$ is the reference epoch, and $P$ is the orbital period. 

The second ephemeris model is a quadratic model which assumes a constant orbital period derivative,
\begin{eqnarray}
T_N &=& T_{N-1}+P_{N-1}, \\
P_N &=& P_{N-1}+\dot{P}~(P_{N-1}+P_N)/2,
\end{eqnarray}
where $P_N$ is the orbital period of the $N$-th epoch. 
Although $\dot{P}$ may evolve over long timescales, it can be treated as constant over the relatively short observational baseline considered in this work, an assumption commonly adopted in studies of tidal orbital decay.

To determine which model better describes the data, we compare their Bayesian information criterion (BIC) values, defined as
\begin{equation}
\mathrm{BIC} = \chi^2 + k \log n,
\end{equation}
where $k$ is the number of free parameters and $n$ the number of data points. 

We then calculate the difference between their BIC scores $\Delta \mathrm{BIC} = \mathrm{BIC}_{\mathrm{lin}} - \mathrm{BIC}_{\mathrm{quad}}$.
To identify candidates with potential orbital period variations, we adopt the following two criteria: the measured $\dot{P}$ deviates from zero at a significance level of at least $3\sigma$; and the quadratic model is strongly preferred, with $\Delta \mathrm{BIC} > 10$.

\section{Results and discussion} \label{sec:Results}

\subsection{Toward an expanded sample of decaying hot Jupiters}
\label{sec:Results.Number}

\begin{figure*}
\centering
\includegraphics[width=\textwidth]{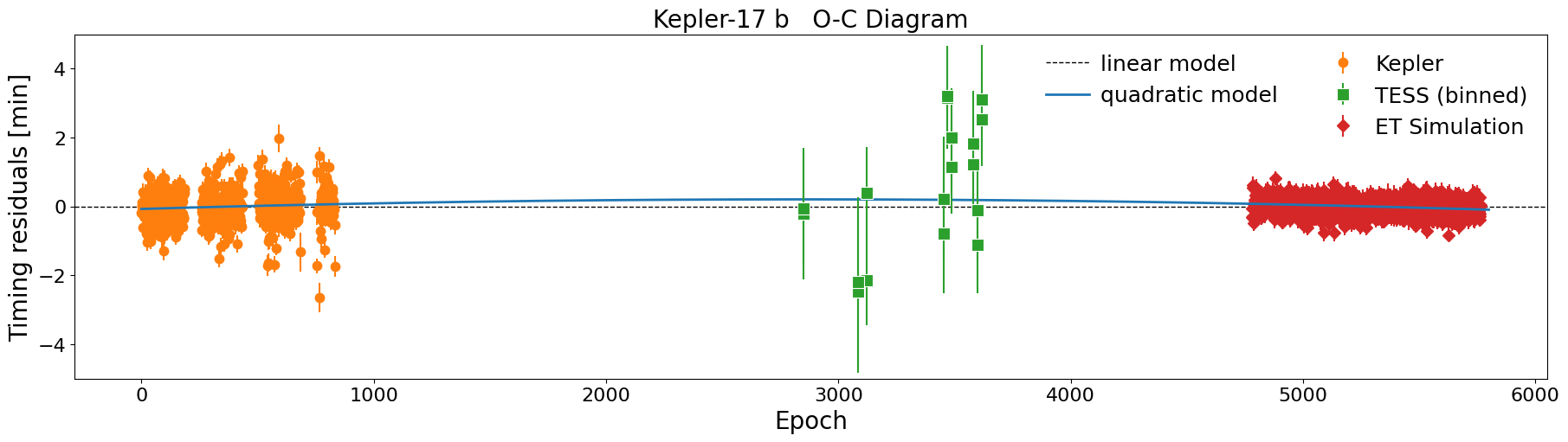}
\caption{
Timing analysis results for Kepler-17 b.
The orange, green, and red points correspond to the \textit{Kepler} (short cadence), \textit{TESS} (sector-binned), and ET-simulated datasets, respectively.
The blue curve represents the best-fit quadratic model, while the horizontal dashed line marks the linear model.
}

\label{fig:kepler17_oc}
\end{figure*}

In this section, we performed the TTV fitting and model comparison analyses for the 28 hot Jupiters by using \texttt{PdotQuest} developed in our previous work \citep{Wang2024}. 
The code fits the transit timing data with both linear and quadratic ephemeris models and compares their BIC values to assess the significance of orbital period variations.
We first fitted each system using only the \textit{Kepler+TESS} mid-transit times.
We then incorporated the simulated ET transit mid-times described in Section~\ref{subsec:ET_Transit} and performed the timing analysis using the combined \textit{Kepler+TESS+ET} dataset. Figure~\ref{fig:kepler17_oc} shows the timing analysis results for Kepler-17~b as an illustrative example of our procedure. The combined \textit{Kepler+TESS+ET} fit yields $\dot{P}=-1.01\pm0.13\,\mathrm{ms\,yr^{-1}}$, with $\Delta\mathrm{BIC}=54.9$, indicating a strong preference for the quadratic model.

Several systems required special treatment due to specific light-curve artifacts.
Kepler-43 b displays an apparent $\approx$ 250-epoch oscillation in transit times derived from the \textit{Kepler} LC data. This signal is caused by aliasing between the planet’s 3.024-day orbital period and the 29.4-min LC sampling, owing to their near-integer ratio. We therefore relied exclusively on the SC \textit{Kepler} data together with TESS observations.
Kepler-63 b and Kepler-75 b exhibit noticeable baseline variability, likely due to stellar rotation and starspot modulation. For Kepler-63 b, the SC timings show larger-than-expected scatter even after detrending; the LC timings were more stable and thus used for the analysis. For Kepler-75 b, several LC transits carry unrealistically small formal uncertainties despite visible scatter; we imposed a minimum uncertainty floor of 0.001 day before fitting.

\begin{figure}
\centering
\includegraphics[width=\hsize]{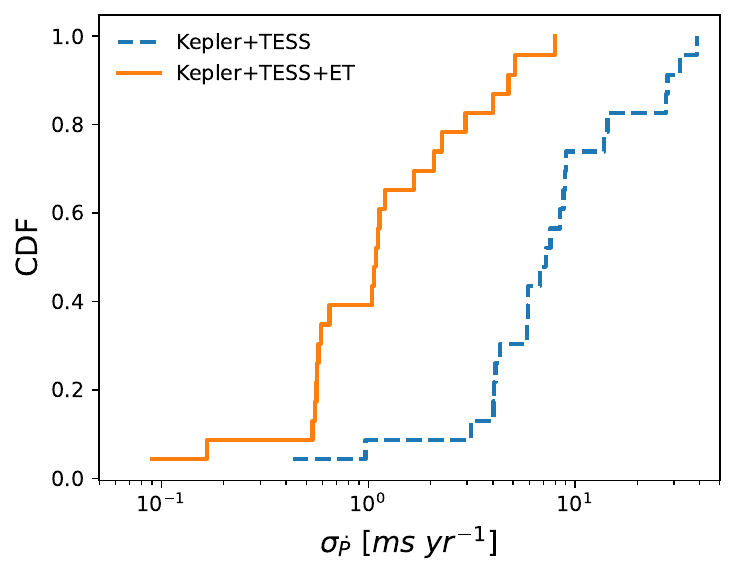}
\caption{
CDFs of the orbital period derivative uncertainties $\sigma_{\dot P}$ for the 23 systems with converged \textit{Kepler+TESS} timing solutions. The blue curve shows the results obtained using only the \textit{Kepler+TESS} mid-transit times, while the orange curve includes the simulated ET transit timings. 
}
\label{fig:epdot_cdf}
\end{figure}

Figure~\ref{fig:epdot_cdf} compares the distributions of the orbital period derivative uncertainties $\sigma_{\dot P}$ obtained from the \textit{Kepler+TESS} and \textit{Kepler+TESS+ET} datasets. 
For the \textit{Kepler}+TESS sample, five systems have transit S/N below the adopted threshold (i.e., ${\rm S/N} < 5$), and therefore do not yield converged $\dot{P}$ measurements. These systems are excluded from Fig.~\ref{fig:epdot_cdf}, leaving 23 systems with converged \textit{Kepler}+TESS solutions for the comparison.
For the \textit{Kepler+TESS} dataset, the typical uncertainty is
$\sigma_{\dot P}=7.2^{+13.7}_{-3.2}\, {\rm ms\,yr^{-1}}$.
After incorporating the simulated ET transit timings, the typical uncertainty decreases to
$\sigma_{\dot P}=1.0^{+3.9}_{-0.6}\,{\rm ms\,yr^{-1}}$.
This corresponds to an improvement of approximately a factor of 7.2 in the median uncertainty,
indicating that ET observations significantly improve the precision of orbital-decay measurements for the hot Jupiter systems considered in this work.

\begin{figure}
\centering
\includegraphics[width=\columnwidth]{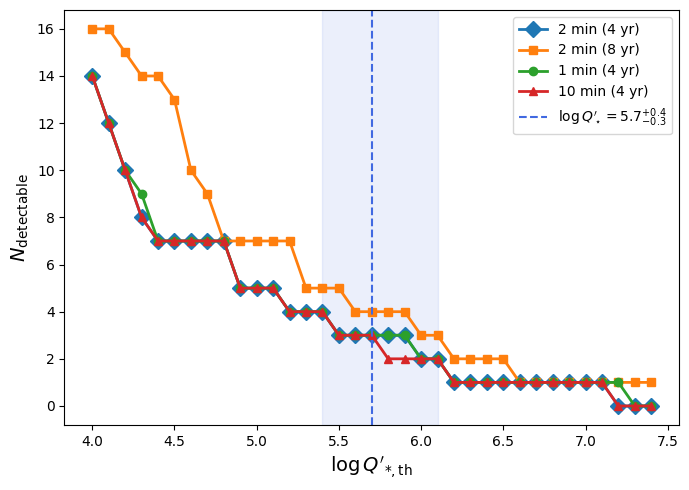}
\caption{
Detectability of tidal orbital decay in the 28 systems considered in this work. The curves show the number of systems satisfying $|\dot{P}_{\mathrm{th}}| > 3\sigma_{\dot{P}}$ as a function of the assumed stellar tidal quality factor, $\log Q'_{\star,\mathrm{th}}$. For each system, the theoretical orbital decay rate $\dot{P}_{\mathrm{th}}$ is computed for a given $Q'_\star$ and compared with the uncertainty in the measured orbital period derivative obtained from joint fits to \textit{Kepler}, TESS, and simulated ET observations. The four curves correspond to ET observing strategies with cadences of 1, 2, and 10 min over a 4 yr baseline, and a 2 min cadence over an 8 yr baseline. The vertical dashed line and shaded region indicate the literature estimate $\log Q'_\star=5.7^{+0.4}_{-0.3}$.
}
\label{fig:Ndetectable_Q}
\end{figure}

We then calculate the theoretically predicted $\dot{P}$ values with Equation \ref{eq:Pdot}.
We then adopt $3\sigma_{\dot{P}}$ as the detection threshold and classify systems exceeding this limit as exhibiting detectable orbital decay.

Figure~\ref{fig:Ndetectable_Q} shows the expected numbers of hot Jupiters with detectable orbital decay as a function of the stellar tidal factor.

After incorporating the simulated
ET transit timings, under the improved precision, three hot Jupiters, i.e., Kepler-13 Ab, Kepler-17 b, and Kepler-76 b, will have absolute orbital decay rates $|\dot{P}|$ exceeding the detection threshold when taking $\log Q^{'}_\star = 5.7^{+0.4}_{-0.3}$ and thus be identified as candidates with statistically significant evidence for orbital decay.

\subsection{Impact of cadence and baseline}

Because the final sampling cadence of the \textit{Earth 2.0} mission has not yet been fixed, we investigated the impact of cadence on the achievable constraints on orbital decay for the full sample. In addition to the nominal 2-minute simulations, we generated ET light curves with cadences of one and ten minutes and repeated the timing analysis for all 28 systems.

For a 4-year observing baseline, the median uncertainty on $\dot{P}$ is
$\sigma_{\dot P}=1.0^{+3.9}_{-0.5}\,\mathrm{ms\,yr^{-1}}$ for the one-minute cadence and
$\sigma_{\dot P}=1.0^{+4.1}_{-0.5}\,\mathrm{ms\,yr^{-1}}$ for the ten-minute cadence.
Thus, the typical uncertainty changes only marginally when the cadence is degraded from one to ten minutes. This is qualitatively consistent with \citet{Price2014}, who showed that finite integration mainly affects transit-timing precision by smoothing the ingress and egress, with the impact depending on the integration time relative to the ingress/egress timescale. This weak dependence on cadence is also reflected in Fig.~\ref{fig:Ndetectable_Q}, where the numbers of detectable systems for the 1, 2, and 10 min cadence cases with a 4-year baseline differ by at most one system.

We also explored extending the ET observing baseline from four to eight years while keeping the nominal 2-minute cadence. In this case, the median uncertainty decreases from
$1.0^{+3.9}_{-0.6}\,\mathrm{ms\,yr^{-1}}$
to
$0.4^{+1.5}_{-0.3}\,\mathrm{ms\,yr^{-1}}$,
corresponding to an improvement of approximately a factor of 2.4 in the median uncertainty.
The stronger impact of extending the observing baseline is also evident in Fig.~\ref{fig:Ndetectable_Q}, where the 2-min, 8-year strategy generally yields more detectable systems than the 4-year strategies across the different cadences.

This is expected, because the quadratic timing signal increases with the square of the observational baseline, whereas cadence affects only the timing precision of individual transits by changing the number of in-transit data points. Once the photometric noise is sufficiently low—as expected for ET—moderate changes in cadence have only a minor impact.

From a mission-planning perspective, this suggests that cadences of one to ten minutes are generally adequate for the detection of tidal orbital decay, providing flexibility for balancing ET’s data volume and operational constraints. In contrast, extending the temporal baseline offers larger scientific returns, either through a longer primary mission or through a potential future revisit of the \textit{Kepler} field. More broadly, these trends reinforce a principle relevant to all transit missions targeting long-term TTVs: maximizing time coverage is far more critical than optimizing cadence when searching for slow secular variations such as tidal decay or LTTE signals.

\section{Conclusion} \label{sec:Conclusion}

In this work, we investigate the capability of the upcoming \textit{Earth 2.0}  mission to improve measurements of orbital decay in hot Jupiter systems by extending the temporal baseline of high-precision transit observations. By combining archival \textit{Kepler} and TESS transit timing measurements with simulated ET observations, we quantify the improvement in the precision of orbital period variation measurements and assess the prospects for detecting tidally driven orbital decay.

We construct a sample of 28 \textit{Kepler} hot Jupiters and extract their mid-transit times from archival \textit{Kepler} and TESS observations. We then generate four years of simulated ET light curves based on the expected photometric performance and observing cadence of the mission, and perform joint long-term transit timing analyses using the combined datasets. We find that the inclusion of ET observations significantly improves the constraints on orbital decay, reducing the uncertainty in the period derivative from $7.2^{+13.7}_{-3.2}\,{\rm ms\,yr^{-1}}$ to $1.0^{+3.9}_{-0.6}\,{\rm ms\,yr^{-1}}$, corresponding to an improvement of approximately a factor of 7.2 across the sample.

Assuming a stellar tidal quality factor of $\log Q^{'}_\star = 5.7^{+0.4}_{-0.3}$ inferred from \citet{Chen2025}, we evaluate the detectability of tidal orbital decay for the 28 hot Jupiter systems. We find that three systems, Kepler-13~Ab, Kepler-17~b, and Kepler-76~b, are expected to exhibit detectable orbital decay with the extended baseline and improved timing precision provided by ET observations.

These results demonstrate that ET will significantly enhance the ability to measure long-term orbital evolution in close-in giant planets. The increased sample of hot Jupiters with detected orbital decay will constrain stellar tidal dissipation efficiency and improve our understanding of the tidal evolution of star-hot Jupiter systems.

\section*{Data availability}
Tables~\ref{tab:sample}--\ref{tab:timing} are available in electronic form at the CDS via anonymous ftp to \texttt{cdsarc.u-strasbg.fr} (130.79.128.5) or via \url{http://cdsweb.u-strasbg.fr/cgi-bin/qcat?J/A+A/}. The full version of Table~\ref{tab:midtc} is available only in electronic form. In the CDS dataset, Tables~\ref{tab:sample} and~\ref{tab:ephemerides} are combined into one table, and Tables~\ref{tab:transitparams} and~\ref{tab:timing} into another.

\begin{acknowledgements}
We thank the anonymous referee for the constructive comments that helped improve this manuscript. The work was supported by the National Key R$\&$D Program of China (2024YFA1611801), the science research grants from the China Manned Space Project (No. CMS-CSST2025-A16), and the \textit{Earth 2.0} Space Mission program from SHAO for their financial support. The ET mission is funded by China's Space Origins Exploration Program. This paper includes data collected by the \textit{Kepler} and \textit{TESS} missions, which are funded by NASA's Science Mission Directorate. We also thank the developers of the open-source software packages \texttt{PyTransit} and \texttt{lightkurve} for their contributions to the scientific community.
\end{acknowledgements}

\bibliographystyle{aa}  
\bibliography{aa62187-26}   

@ARTICLE{Bouma2020,
       author = {{Bouma}, L. G. and {Winn}, J. N. and {Howard}, A. W. and {Howell}, S. B. and {Isaacson}, H. and {Knutson}, H. and {Matson}, R. A.},
        title = {{WASP-4 Is Accelerating toward the Earth}},
      journal = {\apjl},
         year = {2020},
       volume = {893},
        pages = {L29},
          doi = {10.3847/2041-8213/ab8563},
       adsurl = {https://ui.adsabs.harvard.edu/abs/2020ApJ...893L..29B}
}

@ARTICLE{Dawson2018,
       author = {{Dawson}, Rebekah I. and {Johnson}, John Asher},
        title = "{Origins of Hot Jupiters}",
      journal = {\araa},
         year = 2018,
        month = sep,
       volume = {56},
        pages = {175-221},
          doi = {10.1146/annurev-astro-081817-051853},
archivePrefix = {arXiv},
       eprint = {1801.06117},
 primaryClass = {astro-ph.EP},
       adsurl = {https://ui.adsabs.harvard.edu/abs/2018ARA&A..56..175D}
}

@ARTICLE{Rasio1996,
       author = {{Rasio}, F.~A. and {Tout}, C.~A. and {Lubow}, S.~H. and {Livio}, M.},
        title = "{Tidal Decay of Close Planetary Orbits}",
      journal = {\apj},
         year = 1996,
        month = oct,
       volume = {470},
        pages = {1187},
          doi = {10.1086/177941},
archivePrefix = {arXiv},
       eprint = {astro-ph/9605059},
 primaryClass = {astro-ph},
       adsurl = {https://ui.adsabs.harvard.edu/abs/1996ApJ...470.1187R}
}

@ARTICLE{Levrard2009,
       author = {{Levrard}, B. and {Winisdoerffer}, C. and {Chabrier}, G.},
        title = "{Falling Transiting Extrasolar Giant Planets}",
      journal = {\apjl},
         year = 2009,
        month = feb,
       volume = {692},
       number = {1},
        pages = {L9-L13},
          doi = {10.1088/0004-637X/692/1/L9},
archivePrefix = {arXiv},
       eprint = {0901.2048},
 primaryClass = {astro-ph.EP},
       adsurl = {https://ui.adsabs.harvard.edu/abs/2009ApJ...692L...9L}
}

@ARTICLE{Ogilvie2014,
       author = {{Ogilvie}, Gordon I.},
        title = "{Tidal Dissipation in Stars and Giant Planets}",
      journal = {\araa},
         year = 2014,
        month = aug,
       volume = {52},
        pages = {171-210},
          doi = {10.1146/annurev-astro-081913-035941},
archivePrefix = {arXiv},
       eprint = {1406.2207},
 primaryClass = {astro-ph.SR},
       adsurl = {https://ui.adsabs.harvard.edu/abs/2014ARA&A..52..171O}
}

@ARTICLE{Weinberg2017,
       author = {{Weinberg}, Nevin N. and {Sun}, Meng and {Arras}, Phil and {Essick}, Reed},
        title = "{Tidal Dissipation in WASP-12}",
      journal = {\apjl},
         year = 2017,
        month = nov,
       volume = {849},
       number = {1},
          eid = {L11},
        pages = {L11},
          doi = {10.3847/2041-8213/aa9113},
archivePrefix = {arXiv},
       eprint = {1710.00858},
 primaryClass = {astro-ph.EP},
       adsurl = {https://ui.adsabs.harvard.edu/abs/2017ApJ...849L..11W}
}

@ARTICLE{Zahn1977,
       author = {{Zahn}, J.-P.},
        title = "{Tidal friction in close binary systems.}",
      journal = {\aap},
         year = 1977,
        month = may,
       volume = {57},
        pages = {383-394},
       adsurl = {https://ui.adsabs.harvard.edu/abs/1977A&A....57..383Z}
}

@ARTICLE{Zahn1989,
       author = {{Zahn}, J.-P.},
        title = "{Tidal evolution of close binary stars. I - Revisiting the theory of the equilibrium tide}",
      journal = {\aap},
         year = 1989,
        month = aug,
       volume = {220},
       number = {1-2},
        pages = {112-116},
       adsurl = {https://ui.adsabs.harvard.edu/abs/1989A&A...220..112Z}
}

@ARTICLE{Penev2018,
       author = {{Penev}, Kaloyan and {Bouma}, L.~G. and {Winn}, Joshua N. and {Hartman}, Joel D.},
        title = "{Empirical Tidal Dissipation in Exoplanet Hosts From Tidal Spin-up}",
      journal = {\aj},
         year = 2018,
        month = apr,
       volume = {155},
       number = {4},
          eid = {165},
        pages = {165},
          doi = {10.3847/1538-3881/aaaf71},
archivePrefix = {arXiv},
       eprint = {1802.05269},
 primaryClass = {astro-ph.SR},
       adsurl = {https://ui.adsabs.harvard.edu/abs/2018AJ....155..165P}
}

@ARTICLE{Chen2025,
       author = {{Chen}, Di-Chang and {Xie}, Ji-Wei and {Zhou}, Ji-Lin and {Dai}, Fei and {Ma}, Bo and {Wang}, Songhu and {Liu}, Chao},
        title = "{The origin and tidal evolution of hot Jupiters constrained by a broken age-frequency relation}",
      journal = {Nature Astronomy},
         year = 2026,
        month = jan,
       volume = {10},
        pages = {92-104},
          doi = {10.1038/s41550-025-02693-6},
archivePrefix = {arXiv},
       eprint = {2510.26084},
 primaryClass = {astro-ph.EP},
       adsurl = {https://ui.adsabs.harvard.edu/abs/2026NatAs..10...92C}
}

@ARTICLE{Hansen2010,
       author = {{Hansen}, Brad M.~S.},
        title = "{Calibration of Equilibrium Tide Theory for Extrasolar Planet Systems}",
      journal = {\apj},
         year = 2010,
        month = nov,
       volume = {723},
       number = {1},
        pages = {285-299},
          doi = {10.1088/0004-637X/723/1/285},
archivePrefix = {arXiv},
       eprint = {1009.3027},
 primaryClass = {astro-ph.SR},
       adsurl = {https://ui.adsabs.harvard.edu/abs/2010ApJ...723..285H}
}

@ARTICLE{Wang2024,
       author = {{Wang}, Wenqin and {Zhang}, Zixin and {Chen}, Zhangliang and {Wang}, Yonghao and {Yu}, Cong and {Ma}, Bo},
        title = "{Long-term Variations in the Orbital Period of Hot Jupiters from Transit-timing Analysis Using TESS Survey Data}",
      journal = {\apjs},
         year = 2024,
        month = jan,
       volume = {270},
       number = {1},
          eid = {14},
        pages = {14},
          doi = {10.3847/1538-4365/ad0847},
archivePrefix = {arXiv},
       eprint = {2310.17225},
 primaryClass = {astro-ph.EP},
       adsurl = {https://ui.adsabs.harvard.edu/abs/2024ApJS..270...14W}
}

@ARTICLE{Winn2025,
       author = {{Winn}, Joshua N. and {Stef{\'a}nsson}, Gu{\dh}mundur},
        title = "{Orbital decay candidates reconsidered: WASP-4 b is not decaying and Kepler-1658 b is not a planet}",
      journal = {The Planetary Science Journal},
         year = 2025,
        month = dec,
          volume = {6},
       number = {12},
          eid = {300},
        pages = {300},
          doi = {10.3847/PSJ/ae21db},
archivePrefix = {arXiv},
       eprint = {2510.05229},
 primaryClass = {astro-ph.EP},
       adsurl = {https://ui.adsabs.harvard.edu/abs/2025PSJ.....6..300W}
}

@ARTICLE{Maciejewski2016,
       author = {{Maciejewski}, G. and {Dimitrov}, D. and {Fern{\'a}ndez}, M. and {Sota}, A. and {Nowak}, G. and {Ohlert}, J. and {Nikolov}, G. and {Bukowiecki}, {\L}. and {Hinse}, T.~C. and {Pall{\'e}}, E. and {Tingley}, B. and {Kjurkchieva}, D. and {Lee}, J.~W. and {Lee}, C.-U.},
        title = "{Departure from the constant-period ephemeris for the transiting exoplanet WASP-12}",
      journal = {\aap},
         year = 2016,
        month = apr,
       volume = {588},
          eid = {L6},
        pages = {L6},
          doi = {10.1051/0004-6361/201628312},
archivePrefix = {arXiv},
       eprint = {1602.09055},
 primaryClass = {astro-ph.EP},
       adsurl = {https://ui.adsabs.harvard.edu/abs/2016A&A...588L...6M}
}

@ARTICLE{Patra2017,
       author = {{Patra}, Kishore C. and {Winn}, Joshua N. and {Holman}, Matthew J. and {Yu}, Liang and {Deming}, Drake and {Dai}, Fei},
        title = "{The Apparently Decaying Orbit of WASP-12b}",
      journal = {\aj},
         year = 2017,
        month = jul,
       volume = {154},
       number = {1},
          eid = {4},
        pages = {4},
          doi = {10.3847/1538-3881/aa6d75},
archivePrefix = {arXiv},
       eprint = {1703.06582},
 primaryClass = {astro-ph.EP},
       adsurl = {https://ui.adsabs.harvard.edu/abs/2017AJ....154....4P}
}

@ARTICLE{Yee2020,
       author = {{Yee}, Samuel W. and {Winn}, Joshua N. and {Knutson}, Heather A. and {Patra}, Kishore C. and {Vissapragada}, Shreyas and {Zhang}, Michael M. and {Holman}, Matthew J. and {Shporer}, Avi and {Wright}, Jason T.},
        title = "{The Orbit of WASP-12b Is Decaying}",
      journal = {\apjl},
         year = 2020,
        month = jan,
       volume = {888},
       number = {1},
          eid = {L5},
        pages = {L5},
          doi = {10.3847/2041-8213/ab5c16},
archivePrefix = {arXiv},
       eprint = {1911.09131},
 primaryClass = {astro-ph.EP},
       adsurl = {https://ui.adsabs.harvard.edu/abs/2020ApJ...888L...5Y}
}

@ARTICLE{Turner2021,
       author = {{Turner}, Jake D. and {Ridden-Harper}, Andrew and {Jayawardhana}, Ray},
        title = "{Decaying Orbit of the Hot Jupiter WASP-12b: Confirmation with TESS Observations}",
      journal = {\aj},
         year = 2021,
        month = feb,
       volume = {161},
       number = {2},
          eid = {72},
        pages = {72},
          doi = {10.3847/1538-3881/abd178},
archivePrefix = {arXiv},
       eprint = {2012.02211},
 primaryClass = {astro-ph.EP},
       adsurl = {https://ui.adsabs.harvard.edu/abs/2021AJ....161...72T}
}

@ARTICLE{Borucki2010,
       author = {{Borucki}, William J. and {Koch}, David and {Basri}, Gibor and {Batalha}, Natalie and {Brown}, Timothy and {Caldwell}, Douglas and {Caldwell}, John and {Christensen-Dalsgaard}, J{\o}rgen and {Cochran}, William D. and {DeVore}, Edna and {Dunham}, Edward W. and {Dupree}, Andrea K. and {Gautier}, Thomas N. and {Geary}, John C. and {Gilliland}, Ronald and {Gould}, Alan and {Howell}, Steve B. and {Jenkins}, Jon M. and {Kondo}, Yoji and {Latham}, David W. and {Marcy}, Geoffrey W. and {Meibom}, S{\o}ren and {Kjeldsen}, Hans and {Lissauer}, Jack J. and {Monet}, David G. and {Morrison}, David and {Sasselov}, Dimitar and {Tarter}, Jill and {Boss}, Alan and {Brownlee}, Don and {Owen}, Toby and {Buzasi}, Derek and {Charbonneau}, David and {Doyle}, Laurance and {Fortney}, Jonathan and {Ford}, Eric B. and {Holman}, Matthew J. and {Seager}, Sara and {Steffen}, Jason H. and {Welsh}, William F. and {Rowe}, Jason and {Anderson}, Howard and {Buchhave}, Lars and {Ciardi}, David and {Walkowicz}, Lucianne and {Sherry}, William and {Horch}, Elliott and {Isaacson}, Howard and {Everett}, Mark E. and {Fischer}, Debra and {Torres}, Guillermo and {Johnson}, John Asher and {Endl}, Michael and {MacQueen}, Phillip and {Bryson}, Stephen T. and {Dotson}, Jessie and {Haas}, Michael and {Kolodziejczak}, Jeffrey and {Van Cleve}, Jeffrey and {Chandrasekaran}, Hema and {Twicken}, Joseph D. and {Quintana}, Elisa V. and {Clarke}, Bruce D. and {Allen}, Christopher and {Li}, Jie and {Wu}, Haley and {Tenenbaum}, Peter and {Verner}, Ekaterina and {Bruhweiler}, Frederick and {Barnes}, Jason and {Prsa}, Andrej},
        title = "{Kepler Planet-Detection Mission: Introduction and First Results}",
      journal = {Science},
         year = 2010,
        month = feb,
       volume = {327},
       number = {5968},
        pages = {977},
          doi = {10.1126/science.1185402},
       adsurl = {https://ui.adsabs.harvard.edu/abs/2010Sci...327..977B}
}

@ARTICLE{Ricker2015,
       author = {{Ricker}, George R. and {Winn}, Joshua N. and {Vanderspek}, Roland and {Latham}, David W. and {Bakos}, G{\'a}sp{\'a}r {\'A}. and {Bean}, Jacob L. and {Berta-Thompson}, Zachory K. and {Brown}, Timothy M. and {Buchhave}, Lars and {Butler}, Nathaniel R. and {Butler}, R. Paul and {Chaplin}, William J. and {Charbonneau}, David and {Christensen-Dalsgaard}, J{\o}rgen and {Clampin}, Mark and {Deming}, Drake and {Doty}, John and {De Lee}, Nathan and {Dressing}, Courtney and {Dunham}, Edward W. and {Endl}, Michael and {Fressin}, Francois and {Ge}, Jian and {Henning}, Thomas and {Holman}, Matthew J. and {Howard}, Andrew W. and {Ida}, Shigeru and {Jenkins}, Jon M. and {Jernigan}, Garrett and {Johnson}, John Asher and {Kaltenegger}, Lisa and {Kawai}, Nobuyuki and {Kjeldsen}, Hans and {Laughlin}, Gregory and {Levine}, Alan M. and {Lin}, Douglas and {Lissauer}, Jack J. and {MacQueen}, Phillip and {Marcy}, Geoffrey and {McCullough}, Peter R. and {Morton}, Timothy D. and {Narita}, Norio and {Paegert}, Martin and {Palle}, Enric and {Pepe}, Francesco and {Pepper}, Joshua and {Quirrenbach}, Andreas and {Rinehart}, Stephen A. and {Sasselov}, Dimitar and {Sato}, Bun'ei and {Seager}, Sara and {Sozzetti}, Alessandro and {Stassun}, Keivan G. and {Sullivan}, Peter and {Szentgyorgyi}, Andrew and {Torres}, Guillermo and {Udry}, Stephane and {Villasenor}, Joel},
        title = "{Transiting Exoplanet Survey Satellite (TESS)}",
      journal = {Journal of Astronomical Telescopes, Instruments, and Systems},
         year = 2015,
        month = jan,
       volume = {1},
          eid = {014003},
        pages = {014003},
          doi = {10.1117/1.JATIS.1.1.014003},
       adsurl = {https://ui.adsabs.harvard.edu/abs/2015JATIS...1a4003R}
}

@ARTICLE{Patra2020,
       author = {{Patra}, Kishore C. and {Winn}, Joshua N. and {Holman}, Matthew J. and {Gillon}, Michael and {Burdanov}, Artem and {Jehin}, Emmanuel and {Delrez}, Laetitia and {Pozuelos}, Francisco J. and {Barkaoui}, Khalid and {Benkhaldoun}, Zouhair and {Narita}, Norio and {Fukui}, Akihiko and {Kusakabe}, Nobuhiko and {Kawauchi}, Kiyoe and {Terada}, Yuka and {Bouma}, L.~G. and {Weinberg}, Nevin N. and {Broome}, Madelyn},
        title = "{The Continuing Search for Evidence of Tidal Orbital Decay of Hot Jupiters}",
      journal = {\aj},
         year = 2020,
        month = apr,
       volume = {159},
       number = {4},
          eid = {150},
        pages = {150},
          doi = {10.3847/1538-3881/ab7374},
archivePrefix = {arXiv},
       eprint = {2002.02606},
 primaryClass = {astro-ph.EP},
       adsurl = {https://ui.adsabs.harvard.edu/abs/2020AJ....159..150P}
}

@ARTICLE{Ge2022a,
       author = {{Ge}, Jian and {Zhang}, Hui and {the ET Team}},
        title = "{The ET mission to search for earth 2.0s}",
      journal = {The Innovation},
         year = 2022,
        month = jul,
       volume = {3},
          eid = {100271},
        pages = {100271},
          doi = {10.1016/j.xinn.2022.100271},
       adsurl = {https://ui.adsabs.harvard.edu/abs/2022Innov...300271G}
}

@misc{Ge2022b,
	title = {{ET} {White} {Paper}: {To} {Find} the {First} {Earth} 2.0},
	shorttitle = {{ET} {White} {Paper}},
	url = {http://arxiv.org/abs/2206.06693},
	doi = {10.48550/arXiv.2206.06693},
	urldate = {2025-11-18},
	publisher = {arXiv},
	author = {{Ge}, Jian and {Zhang}, Hui and {the ET Team}},
	month = jun,
	year = {2022},
	note = {arXiv:2206.06693 [astro-ph]},
}

@ARTICLE{Ge2024,
       author = {{Ge}, Jian and {Chen}, Wen and {Chen}, Yonghe and {Song}, Zongxi and {Wang}, Jian and {Zhang}, Hui and {Li}, Yan and {Zang}, Weicheng and {Zhou}, Dan and {Zhang}, Yongshuai and {Chen}, Kun and {Yang}, Yingquan and {Mao}, Shude and {Huang}, Chelsea and {Yao}, Xinyu and {Li}, Xinglong and {Jiang}, Haijiao and {Yu}, Yong and {Tang}, Zhenghong and {Dong}, Feng and {Gao}, Wei and {Zhang}, Hongfei and {Shen}, Chao and {Wang}, Fengtao and {Wei}, Chuanxin and {Yang}, Baoyu and {Li}, Yudong and {Wen}, Lin and {Zhang}, Pengjun and {Zhang}, Congcong and {Xie}, Jiwei and {Ma}, Bo and {Deng}, Hongping and {Liu}, Huigen and {Duan}, Xuliang and {Wang}, Haoyu and {Huang}, Jiangjiang and {Gao}, Yang and {Wang}, Yifei and {Wang}, Lei and {Qin}, Genjian and {Liu}, Xinyu and {Gao}, Jie},
        title = "{Search for a Second Earth - the Earth 2.0 (ET) Space Mission}",
      journal = {Chinese Journal of Space Science},
         year = 2024,
        month = may,
       volume = {44},
       number = {3},
        pages = {400-424},
          doi = {10.11728/cjss2024.03.yg05},
       adsurl = {https://ui.adsabs.harvard.edu/abs/2024ChJSS..44..400G}
}

@ARTICLE{Lillo-Box2014L,
       author = {{Lillo-Box}, J. and {Barrado}, D. and {Moya}, A. and {Montesinos}, B. and {Montalb{\'a}n}, J. and {Bayo}, A. and {Barbieri}, M. and {R{\'e}gulo}, C. and {Mancini}, L. and {Bouy}, H. and {Henning}, T.},
        title = "{Kepler-91b: a planet at the end of its life. Planet and giant host star properties via light-curve variations}",
      journal = {\aap},
         year = 2014,
        month = feb,
       volume = {562},
          eid = {A109},
        pages = {A109},
          doi = {10.1051/0004-6361/201322001},
archivePrefix = {arXiv},
       eprint = {1312.3943},
 primaryClass = {astro-ph.EP},
       adsurl = {https://ui.adsabs.harvard.edu/abs/2014A&A...562A.109L}
}

@ARTICLE{Mandel2002,
       author = {{Mandel}, Kaisey and {Agol}, Eric},
        title = "{Analytic Light Curves for Planetary Transit Searches}",
      journal = {\apjl},
         year = 2002,
        month = dec,
       volume = {580},
       number = {2},
        pages = {L171-L175},
          doi = {10.1086/345520},
archivePrefix = {arXiv},
       eprint = {astro-ph/0210099},
 primaryClass = {astro-ph},
       adsurl = {https://ui.adsabs.harvard.edu/abs/2002ApJ...580L.171M}
}

@ARTICLE{Goldreich1966,
       author = {{Goldreich}, Peter and {Soter}, Steven},
        title = "{Q in the Solar System}",
      journal = {\icarus},
         year = 1966,
        month = jan,
       volume = {5},
       number = {1},
        pages = {375-389},
          doi = {10.1016/0019-1035(66)90051-0},
       adsurl = {https://ui.adsabs.harvard.edu/abs/1966Icar....5..375G}
}

@ARTICLE{Esteves2015,
       author = {{Esteves}, Lisa J. and {De Mooij}, Ernst J.~W. and {Jayawardhana}, Ray},
        title = "{Changing Phases of Alien Worlds: Probing Atmospheres of Kepler Planets with High-precision Photometry}",
      journal = {\apj},
         year = 2015,
        month = may,
       volume = {804},
       number = {2},
          eid = {150},
        pages = {150},
          doi = {10.1088/0004-637X/804/2/150},
archivePrefix = {arXiv},
       eprint = {1407.2245},
 primaryClass = {astro-ph.EP},
       adsurl = {https://ui.adsabs.harvard.edu/abs/2015ApJ...804..150E}
}

@ARTICLE{Endl2011,
       author = {{Endl}, Michael and {MacQueen}, Phillip J. and {Cochran}, William D. and {Brugamyer}, Erik J. and {Buchhave}, Lars A. and {Rowe}, Jason and {Lucas}, Phillip and {Isaacson}, Howard and {Bryson}, Steve and {Howell}, Steve B. and {Fortney}, Jonathan J. and {Hansen}, Terese and {Borucki}, William J. and {Caldwell}, Douglas and {Christiansen}, Jessie L. and {Ciardi}, David R. and {Demory}, Brice-Olivier and {Everett}, Mark and {Ford}, Eric B. and {Haas}, Michael R. and {Holman}, Matthew J. and {Horch}, Elliott and {Jenkins}, Jon M. and {Koch}, David J. and {Lissauer}, Jack J. and {Machalek}, Pavel and {Still}, Martin and {Welsh}, William F. and {Sanderfer}, Dwight T. and {Seader}, Shawn E. and {Smith}, Jeffrey C. and {Thompson}, Susan E. and {Twicken}, Joseph D.},
        title = "{Kepler-15b: A Hot Jupiter Enriched in Heavy Elements and the First Kepler Mission Planet Confirmed with the Hobby-Eberly Telescope}",
      journal = {\apjs},
         year = 2011,
        month = nov,
       volume = {197},
       number = {1},
          eid = {13},
        pages = {13},
          doi = {10.1088/0067-0049/197/1/13},
       adsurl = {https://ui.adsabs.harvard.edu/abs/2011ApJS..197...13E}
}

@ARTICLE{Almenara2015,
       author = {{Almenara}, J.~M. and {Damiani}, C. and {Bouchy}, F. and {Havel}, M. and {Bruno}, G. and {H{\'e}brard}, G. and {Diaz}, R.~F. and {Deleuil}, M. and {Barros}, S.~C.~C. and {Boisse}, I. and {Bonomo}, A.~S. and {Montagnier}, G. and {Santerne}, A.},
        title = "{SOPHIE velocimetry of Kepler transit candidates. XV. KOI-614b, KOI-206b, and KOI-680b: a massive warm Jupiter orbiting a G0 metallic dwarf and two highly inflated planets with a distant companion around evolved F-type stars}",
      journal = {\aap},
         year = 2015,
        month = mar,
       volume = {575},
          eid = {A71},
        pages = {A71},
          doi = {10.1051/0004-6361/201424291},
archivePrefix = {arXiv},
       eprint = {1501.01486},
 primaryClass = {astro-ph.EP},
       adsurl = {https://ui.adsabs.harvard.edu/abs/2015A&A...575A..71A}
}

@ARTICLE{Buchhave2011,
       author = {{Buchhave}, Lars A. and {Latham}, David W. and {Carter}, Joshua A. and {D{\'e}sert}, Jean-Michel and {Torres}, Guillermo and {Adams}, Elisabeth R. and {Bryson}, Stephen T. and {Charbonneau}, David B. and {Ciardi}, David R. and {Kulesa}, Craig and {Dupree}, Andrea K. and {Fischer}, Debra A. and {Fressin}, Fran{\c{c}}ois and {Gautier}, III, Thomas N. and {Gilliland}, Ronald L. and {Howell}, Steve B. and {Isaacson}, Howard and {Jenkins}, Jon M. and {Marcy}, Geoffrey W. and {McCarthy}, Donald W. and {Rowe}, Jason F. and {Batalha}, Natalie M. and {Borucki}, William J. and {Brown}, Timothy M. and {Caldwell}, Douglas A. and {Christiansen}, Jessie L. and {Cochran}, William D. and {Deming}, Drake and {Dunham}, Edward W. and {Everett}, Mark and {Ford}, Eric B. and {Fortney}, Jonathan J. and {Geary}, John C. and {Girouard}, Forrest R. and {Haas}, Michael R. and {Holman}, Matthew J. and {Horch}, Elliott and {Klaus}, Todd C. and {Knutson}, Heather A. and {Koch}, David G. and {Kolodziejczak}, Jeffrey and {Lissauer}, Jack J. and {Machalek}, Pavel and {Mullally}, Fergal and {Still}, Martin D. and {Quinn}, Samuel N. and {Seager}, Sara and {Thompson}, Susan E. and {Van Cleve}, Jeffrey},
        title = "{Kepler-14b: A Massive Hot Jupiter Transiting an F Star in a Close Visual Binary}",
      journal = {\apjs},
         year = 2011,
        month = nov,
       volume = {197},
       number = {1},
          eid = {3},
        pages = {3},
          doi = {10.1088/0067-0049/197/1/3},
archivePrefix = {arXiv},
       eprint = {1106.5510},
 primaryClass = {astro-ph.EP},
       adsurl = {https://ui.adsabs.harvard.edu/abs/2011ApJS..197....3B}
}

@ARTICLE{Desert2011,
       author = {{D{\'e}sert}, Jean-Michel and {Charbonneau}, David and {Demory}, Brice-Olivier and {Ballard}, Sarah and {Carter}, Joshua A. and {Fortney}, Jonathan J. and {Cochran}, William D. and {Endl}, Michael and {Quinn}, Samuel N. and {Isaacson}, Howard T. and {Fressin}, Fran{\c{c}}ois and {Buchhave}, Lars A. and {Latham}, David W. and {Knutson}, Heather A. and {Bryson}, Stephen T. and {Torres}, Guillermo and {Rowe}, Jason F. and {Batalha}, Natalie M. and {Borucki}, William J. and {Brown}, Timothy M. and {Caldwell}, Douglas A. and {Christiansen}, Jessie L. and {Deming}, Drake and {Fabrycky}, Daniel C. and {Ford}, Eric B. and {Gilliland}, Ronald L. and {Gillon}, Micha{\"e}l and {Haas}, Micha{\"e}l R. and {Jenkins}, Jon M. and {Kinemuchi}, Karen and {Koch}, David and {Lissauer}, Jack J. and {Lucas}, Philip and {Mullally}, Fergal and {MacQueen}, Phillip J. and {Marcy}, Geoffrey W. and {Sasselov}, Dimitar D. and {Seager}, Sara and {Still}, Martin and {Tenenbaum}, Peter and {Uddin}, Kamal and {Winn}, Joshua N.},
        title = "{The Hot-Jupiter Kepler-17b: Discovery, Obliquity from Stroboscopic Starspots, and Atmospheric Characterization}",
      journal = {\apjs},
         year = 2011,
        month = nov,
       volume = {197},
       number = {1},
          eid = {14},
        pages = {14},
          doi = {10.1088/0067-0049/197/1/14},
archivePrefix = {arXiv},
       eprint = {1107.5750},
 primaryClass = {astro-ph.EP},
       adsurl = {https://ui.adsabs.harvard.edu/abs/2011ApJS..197...14D}
}

@ARTICLE{Santerne2011,
       author = {{Santerne}, A. and {D{\'\i}az}, R.~F. and {Bouchy}, F. and {Deleuil}, M. and {Moutou}, C. and {H{\'e}brard}, G. and {Eggenberger}, A. and {Ehrenreich}, D. and {Gry}, C. and {Udry}, S.},
        title = "{SOPHIE velocimetry of Kepler transit candidates. II. KOI-428b: a hot Jupiter transiting a subgiant F-star}",
      journal = {\aap},
         year = 2011,
        month = apr,
       volume = {528},
          eid = {A63},
        pages = {A63},
          doi = {10.1051/0004-6361/201015764},
archivePrefix = {arXiv},
       eprint = {1101.0196},
 primaryClass = {astro-ph.EP},
       adsurl = {https://ui.adsabs.harvard.edu/abs/2011A&A...528A..63S}
}

@ARTICLE{Bonomo2012,
       author = {{Bonomo}, A. S. and {H{\'e}brard}, G. and {Santerne}, A. and {Santos}, N. C. and {Deleuil}, M. and {Almenara}, J. and {Bouchy}, F. and {D{\'i}az}, R. F. and {Moutou}, C. and {Vanhuysse}, M.},
        title = {SOPHIE velocimetry of Kepler transit candidates. V. The three hot Jupiters KOI-135b, KOI-204b, and KOI-203b (alias Kepler-17b)},
      journal = {\aap},
         year = {2012},
       volume = {538},
        pages = {A96},
          doi = {10.1051/0004-6361/201118323},
       adsurl = {https://ui.adsabs.harvard.edu/abs/2012A&A...538A..96B}
}

@ARTICLE{Bonomo2017,
       author = {{Bonomo}, A.~S. and {Desidera}, S. and {Benatti}, S. and {Borsa}, F. and {Crespi}, S. and {Damasso}, M. and {Lanza}, A.~F. and {Sozzetti}, A. and {Lodato}, G. and {Marzari}, F. and {Boccato}, C. and {Claudi}, R.~U. and {Cosentino}, R. and {Covino}, E. and {Gratton}, R. and {Maggio}, A. and {Micela}, G. and {Molinari}, E. and {Pagano}, I. and {Piotto}, G. and {Poretti}, E. and {Smareglia}, R. and {Affer}, L. and {Biazzo}, K. and {Bignamini}, A. and {Esposito}, M. and {Giacobbe}, P. and {H{\'e}brard}, G. and {Malavolta}, L. and {Maldonado}, J. and {Mancini}, L. and {Martinez Fiorenzano}, A. and {Masiero}, S. and {Nascimbeni}, V. and {Pedani}, M. and {Rainer}, M. and {Scandariato}, G.},
        title = "{The GAPS Programme with HARPS-N at TNG . XIV. Investigating giant planet migration history via improved eccentricity and mass determination for 231 transiting planets}",
      journal = {\aap},
         year = 2017,
        month = jun,
       volume = {602},
          eid = {A107},
        pages = {A107},
          doi = {10.1051/0004-6361/201629882},
archivePrefix = {arXiv},
       eprint = {1704.00373},
 primaryClass = {astro-ph.EP},
       adsurl = {https://ui.adsabs.harvard.edu/abs/2017A&A...602A.107B}
}

@ARTICLE{Johnson2012,
       author = {{Johnson}, John Asher and {Gazak}, J. Zachary and {Apps}, Kevin and {Muirhead}, Philip S. and {Crepp}, Justin R. and {Crossfield}, Ian J.~M. and {Boyajian}, Tabetha and {von Braun}, Kaspar and {Rojas-Ayala}, Barbara and {Howard}, Andrew W. and {Covey}, Kevin R. and {Schlawin}, Everett and {Hamren}, Katherine and {Morton}, Timothy D. and {Marcy}, Geoffrey W. and {Lloyd}, James P.},
        title = "{Characterizing the Cool KOIs. II. The M Dwarf KOI-254 and Its Hot Jupiter}",
      journal = {\aj},
         year = 2012,
        month = may,
       volume = {143},
       number = {5},
          eid = {111},
        pages = {111},
          doi = {10.1088/0004-6256/143/5/111},
archivePrefix = {arXiv},
       eprint = {1112.0017},
 primaryClass = {astro-ph.EP},
       adsurl = {https://ui.adsabs.harvard.edu/abs/2012AJ....143..111J}
}

@ARTICLE{Sanchis2013,
       author = {{Sanchis-Ojeda}, Roberto and {Winn}, Joshua N. and {Marcy}, Geoffrey W. and {Howard}, Andrew W. and {Isaacson}, Howard and {Johnson}, John Asher and {Torres}, Guillermo and {Albrecht}, Simon and {Campante}, Tiago L. and {Chaplin}, William J. and {Davies}, Guy R. and {Lund}, Mikkel N. and {Carter}, Joshua A. and {Dawson}, Rebekah I. and {Buchhave}, Lars A. and {Everett}, Mark E. and {Fischer}, Debra A. and {Geary}, John C. and {Gilliland}, Ronald L. and {Horch}, Elliott P. and {Howell}, Steve B. and {Latham}, David W.},
        title = "{Kepler-63b: A Giant Planet in a Polar Orbit around a Young Sun-like Star}",
      journal = {\apj},
         year = 2013,
        month = sep,
       volume = {775},
       number = {1},
          eid = {54},
        pages = {54},
          doi = {10.1088/0004-637X/775/1/54},
archivePrefix = {arXiv},
       eprint = {1307.8128},
 primaryClass = {astro-ph.EP},
       adsurl = {https://ui.adsabs.harvard.edu/abs/2013ApJ...775...54S}
}

@ARTICLE{Gandolfi2013,
       author = {{Gandolfi}, D. and {Parviainen}, H. and {Fridlund}, M. and {Hatzes}, A.~P. and {Deeg}, H.~J. and {Frasca}, A. and {Lanza}, A.~F. and {Prada Moroni}, P.~G. and {Tognelli}, E. and {McQuillan}, A. and {Aigrain}, S. and {Alonso}, R. and {Antoci}, V. and {Cabrera}, J. and {Carone}, L. and {Csizmadia}, Sz. and {Djupvik}, A.~A. and {Guenther}, E.~W. and {Jessen-Hansen}, J. and {Ofir}, A. and {Telting}, J.},
        title = "{Kepler-77b: a very low albedo, Saturn-mass transiting planet around a metal-rich solar-like star}",
      journal = {\aap},
         year = 2013,
        month = sep,
       volume = {557},
          eid = {A74},
        pages = {A74},
          doi = {10.1051/0004-6361/201321901},
archivePrefix = {arXiv},
       eprint = {1305.3891},
 primaryClass = {astro-ph.EP},
       adsurl = {https://ui.adsabs.harvard.edu/abs/2013A&A...557A..74G}
}

@ARTICLE{Endl2014,
       author = {{Endl}, Michael and {Caldwell}, Douglas A. and {Barclay}, Thomas and {Huber}, Daniel and {Isaacson}, Howard and {Buchhave}, Lars A. and {Brugamyer}, Erik and {Robertson}, Paul and {Cochran}, William D. and {MacQueen}, Phillip J. and {Havel}, Mathieu and {Lucas}, Phillip and {Howell}, Steve B. and {Fischer}, Debra and {Quintana}, Elisa and {Ciardi}, David R.},
        title = "{Kepler-424 b: A ``Lonely'' Hot Jupiter that Found a Companion}",
      journal = {\apj},
         year = 2014,
        month = nov,
       volume = {795},
       number = {2},
          eid = {151},
        pages = {151},
          doi = {10.1088/0004-637X/795/2/151},
archivePrefix = {arXiv},
       eprint = {1409.8156},
 primaryClass = {astro-ph.EP},
       adsurl = {https://ui.adsabs.harvard.edu/abs/2014ApJ...795..151E}
}

@ARTICLE{Gandolfi2015,
       author = {{Gandolfi}, D. and {Parviainen}, H. and {Deeg}, H.~J. and {Lanza}, A.~F. and {Fridlund}, M. and {Prada Moroni}, P.~G. and {Alonso}, R. and {Augusteijn}, T. and {Cabrera}, J. and {Evans}, T. and {Geier}, S. and {Hatzes}, A.~P. and {Holczer}, T. and {Hoyer}, S. and {Kangas}, T. and {Mazeh}, T. and {Pagano}, I. and {Tal-Or}, L. and {Tingley}, B.},
        title = "{Kepler-423b: a half-Jupiter mass planet transiting a very old solar-like star}",
      journal = {\aap},
         year = 2015,
        month = apr,
       volume = {576},
          eid = {A11},
        pages = {A11},
          doi = {10.1051/0004-6361/201425062},
archivePrefix = {arXiv},
       eprint = {1409.8245},
 primaryClass = {astro-ph.EP},
       adsurl = {https://ui.adsabs.harvard.edu/abs/2015A&A...576A..11G}
}

@ARTICLE{Hebrard2014,
       author = {{H{\'e}brard}, G. and {Santerne}, A. and {Montagnier}, G. and {Bruno}, G. and {Deleuil}, M. and {Havel}, M. and {Almenara}, J.-M. and {Damiani}, C. and {Barros}, S.~C.~C. and {Bonomo}, A.~S. and {Bouchy}, F. and {D{\'\i}az}, R.~F. and {Moutou}, C.},
        title = "{Characterization of the four new transiting planets KOI-188b, KOI-195b, KOI-192b, and KOI-830b}",
      journal = {\aap},
         year = 2014,
        month = dec,
       volume = {572},
          eid = {A93},
        pages = {A93},
          doi = {10.1051/0004-6361/201424268},
archivePrefix = {arXiv},
       eprint = {1409.8554},
 primaryClass = {astro-ph.EP},
       adsurl = {https://ui.adsabs.harvard.edu/abs/2014A&A...572A..93H}
}

@ARTICLE{Lillo2015,
       author = {{Lillo-Box}, J. and {Barrado}, D. and {Santos}, N.~C. and {Mancini}, L. and {Figueira}, P. and {Ciceri}, S. and {Henning}, Th.},
        title = "{Kepler-447b: a hot-Jupiter with an extremely grazing transit}",
      journal = {\aap},
         year = 2015,
        month = may,
       volume = {577},
          eid = {A105},
        pages = {A105},
          doi = {10.1051/0004-6361/201425428},
archivePrefix = {arXiv},
       eprint = {1502.03267},
 primaryClass = {astro-ph.EP},
       adsurl = {https://ui.adsabs.harvard.edu/abs/2015A&A...577A.105L}
}

@misc{2018ascl.soft12013L,
       author = {{Lightkurve Collaboration} and {Cardoso}, Jos{\'e} Vin{\'\i}cius de Miranda and {Hedges}, Christina and {Gully-Santiago}, Michael and {Saunders}, Nicholas and {Cody}, Ann Marie and {Barclay}, Thomas and {Hall}, Oliver and {Sagear}, Sheila and {Turtelboom}, Emma and {Zhang}, Johnny and {Tzanidakis}, Andy and {Mighell}, Ken and {Coughlin}, Jeff and {Bell}, Keaton and {Berta-Thompson}, Zach and {Williams}, Peter and {Dotson}, Jessie and {Barentsen}, Geert},
        title = "{Lightkurve: Kepler and TESS time series analysis in Python}",
 howpublished = {Astrophysics Source Code Library, record ascl:1812.013},
         year = 2018,
        month = dec,
          eid = {ascl:1812.013},
archivePrefix = {ascl},
       eprint = {1812.013},
       adsurl = {https://ui.adsabs.harvard.edu/abs/2018ascl.soft12013L}
}

@ARTICLE{Parviainen2015pytransit,
       author = {{Parviainen}, Hannu},
        title = "{PYTRANSIT: fast and easy exoplanet transit modelling in PYTHON}",
      journal = {\mnras},
         year = 2015,
        month = jul,
       volume = {450},
       number = {3},
        pages = {3233-3238},
          doi = {10.1093/mnras/stv894},
archivePrefix = {arXiv},
       eprint = {1504.07433},
 primaryClass = {astro-ph.EP},
       adsurl = {https://ui.adsabs.harvard.edu/abs/2015MNRAS.450.3233P}
}

@ARTICLE{Parviainen2015LDTK,
       author = {{Parviainen}, H. and {Aigrain}, S.},
        title = "{LDTK: Limb Darkening Toolkit}",
      journal = {\mnras},
         year = 2015,
        month = nov,
       volume = {453},
       number = {4},
        pages = {3821-3826},
          doi = {10.1093/mnras/stv1857},
archivePrefix = {arXiv},
       eprint = {1508.02634},
 primaryClass = {astro-ph.EP},
       adsurl = {https://ui.adsabs.harvard.edu/abs/2015MNRAS.453.3821P}
}

@ARTICLE{Zhangzixin24,
       author = {{Zhang}, Zixin and {Wang}, Wenqin and {Ma}, Xinyue and {Chen}, Zhangliang and {Wang}, Yonghao and {Yu}, Cong and {Liu}, Shangfei and {Gao}, Yang and {Tang}, Baitian and {Ma}, Bo},
        title = "{Constraining the Presence of Companion Planets in Hot Jupiter Planetary Systems Using Transit-timing Variation Observations from TESS}",
      journal = {\apjs},
         year = 2024,
        month = dec,
       volume = {275},
       number = {2},
          eid = {32},
        pages = {32},
          doi = {10.3847/1538-4365/ad89a6},
archivePrefix = {arXiv},
       eprint = {2410.03101},
 primaryClass = {astro-ph.EP},
       adsurl = {https://ui.adsabs.harvard.edu/abs/2024ApJS..275...32Z}
}

@ARTICLE{Zhangzixin26,
       author = {{Zhang}, Zixin and {Wang}, Wenqin and {Ma}, Xinyue and {Chen}, Zhangliang and {Wang}, Yonghao and {Yu}, Cong and {Liu}, Shangfei and {Gao}, Yang and {Tang}, Baitian and {Chen}, Dichang and {Ma}, Bo},
        title = "{Constraining exotrojans in hot-Jupiter systems using transit timing variation observations from TESS}",
      journal = {\aap},
         year = 2026,
        month = apr,
       volume = {709},
          eid = {A35},
        pages = {A35},
          doi = {10.1051/0004-6361/202557704},
archivePrefix = {arXiv},
       eprint = {2603.27215},
 primaryClass = {astro-ph.EP},
       adsurl = {https://ui.adsabs.harvard.edu/abs/2026A&A...709A..35Z}
}

@ARTICLE{Maxinyue2025,
       author = {{Ma}, Xinyue and {Wang}, Wenqin and {Zhang}, Zixin and {Yu}, Cong and {Chen}, Dichang and {Xie}, Jiwei and {Liu}, Shangfei and {Zhou}, Li and {Ma}, Bo},
        title = "{Exoplanet Ephemerides Change Observations (ExoEcho). I. Transit Timing Analysis of 37 Exoplanets Using HST/WFC3 Data}",
      journal = {\aj},
         year = 2025,
        month = mar,
       volume = {169},
       number = {3},
          eid = {169},
        pages = {169},
          doi = {10.3847/1538-3881/ada95d},
archivePrefix = {arXiv},
       eprint = {2501.05704},
 primaryClass = {astro-ph.EP},
       adsurl = {https://ui.adsabs.harvard.edu/abs/2025AJ....169..169M}
}

@ARTICLE{Wu2026,
       author = {{Wu}, Yunke and {Zhang}, Zixin and {Ma}, Xinyue and {Chen}, Zhangliang and {Wang}, Wenqin and {Liu}, Shangfei and {Yu}, Cong and {Chen}, Dichang and {Ma}, Bo},
        title = "{Evidence for a Nonzero Eccentricity Superpuff Exoplanet WASP-107 b Using JWST Occultation Observation}",
      journal = {\apjl},
         year = 2026,
        month = jan,
       volume = {996},
       number = {2},
          eid = {L28},
        pages = {L28},
          doi = {10.3847/2041-8213/ae2ac8},
archivePrefix = {arXiv},
       eprint = {2601.00640},
 primaryClass = {astro-ph.EP},
       adsurl = {https://ui.adsabs.harvard.edu/abs/2026ApJ...996L..28W}
}

@ARTICLE{Wang2025,
       author = {{Wang}, Wenqin and {Ma}, Xinyue and {Chen}, Zhangliang and {Yu}, Cong and {Liu}, Shangfei and {Ma}, Bo},
        title = "{Exoplanet Ephemerides Change Observations (ExoEcho). II. Transit Timing Variation Analysis of Brown Dwarfs around Solar-type Stars}",
      journal = {\apj},
         year = 2025,
        month = jun,
       volume = {986},
       number = {2},
          eid = {121},
        pages = {121},
          doi = {10.3847/1538-4357/add3f2},
archivePrefix = {arXiv},
       eprint = {2505.21270},
 primaryClass = {astro-ph.EP},
       adsurl = {https://ui.adsabs.harvard.edu/abs/2025ApJ...986..121W}
}

@ARTICLE{Cuizimiao2026,
       author = {{Cui}, Zimiao and {Zhang}, Zixin and {Wang}, Wenqin and {Ma}, Xinyue and {Wu}, Yunke and {Zhang}, Hui and {Ge}, Jian and {Ma}, Bo},
        title = "{A simulation study of detecting exomoons using the Earth 2.0 (ET) mission}",
      journal = {\aap},
         year = 2026,
        month = apr,
       volume = {708},
          eid = {A372},
        pages = {A372},
          doi = {10.1051/0004-6361/202558332},
       adsurl = {https://ui.adsabs.harvard.edu/abs/2026A&A...708A.372C}
}

@ARTICLE{Price2014,
       author = {{Price}, Ellen M. and {Rogers}, Leslie A.},
        title = "{Transit Light Curves with Finite Integration Time: Fisher Information Analysis}",
      journal = {\apj},
         year = 2014,
        month = oct,
       volume = {794},
       number = {1},
          eid = {92},
        pages = {92},
          doi = {10.1088/0004-637X/794/1/92},
archivePrefix = {arXiv},
       eprint = {1408.4124},
 primaryClass = {astro-ph.EP},
       adsurl = {https://ui.adsabs.harvard.edu/abs/2014ApJ...794...92P}
}

@ARTICLE{Bonomo2015,
  author = {{Bonomo}, A. S. and {Sozzetti}, A. and {Santerne}, A. and {Deleuil}, M. and {Almenara}, J.-M. and {Bruno}, G. and {Díaz}, R. F. and {Hébrard}, G. and {Moutou}, C.},
  title = {Improved parameters of seven<i>Kepler</i>giant companions characterized with SOPHIE and HARPS-N},
  journal = {\aap},
  year = {2015},
  volume = {575},
  pages = {A85},
  doi = {10.1051/0004-6361/201323042},
}

@ARTICLE{Kokori2023,
  author = {{Kokori}, A. and {Tsiaras}, A. and {Edwards}, B. and {Jones}, A. and {Pantelidou}, G. and {Tinetti}, G. and {Bewersdorff}, L. and {Iliadou}, A. and {Jongen}, Y. and {Lekkas}, G. and {Nastasi}, A. and {Poultourtzidis}, E. and {Sidiropoulos}, C. and {Walter}, F. and {Wünsche}, A. and {Abraham}, R. and {Agnihotri}, V. K. and {Albanesi}, R. and {Arce-Mansego}, E. and {Arnot}, D. and {Audejean}, M. and {Aumasson}, C. and {Bachschmidt}, M. and {Baj}, G. and {Barroy}, P. R. and {Belinski}, A. A. and {Bennett}, D. and {Benni}, P. and {Bernacki}, K. and {Betti}, L. and {Biagini}, A. and {Bosch}, P. and {Brandebourg}, P. and {Brát}, L. and {Bretton}, M. and {Brincat}, S. M. and {Brouillard}, S. and {Bruzas}, A. and {Bruzzone}, A. and {Buckland}, R. A. and {Caló}, M. and {Campos}, F. and {Carreño}, A. and {Carrion Rodrigo}, J. A. and {Casali}, R. and {Casalnuovo}, G. and {Cataneo}, M. and {Chang}, C.-M. and {Changeat}, L. and {Chowdhury}, V. and {Ciantini}, R. and {Cilluffo}, M. and {Coliac}, J.-F. and {Conzo}, G. and {Correa}, M. and {Coulon}, G. and {Crouzet}, N. and {Crow}, M. V. and {Curtis}, I. A. and {Daniel}, D. and {Dauchet}, B. and {Dawes}, S. and {Deldem}, M. and {Deligeorgopoulos}, D. and {Dransfield}, G. and {Dymock}, R. and {Eenmäe}, T. and {Esseiva}, N. and {Evans}, P. and {Falco}, C. and {Farfán}, R. G. and {Fernández-Lajús}, E. and {Ferratfiat}, S. and {Ferreira}, S. L. and {Ferretti}, A. and {Fiołka}, J. and {Fowler}, M. and {Futcher}, S. R. and {Gabellini}, D. and {Gainey}, T. and {Gaitan}, J. and {Gajdoš}, P. and {García-Sánchez}, A. and {Garlitz}, J. and {Gillier}, C. and {Gison}, C. and {Gonzales}, J. and {Gorshanov}, D. and {Grau Horta}, F. and {Grivas}, G. and {Guerra}, P. and {Guillot}, T. and {Haswell}, C. A. and {Haymes}, T. and {Hentunen}, V.-P. and {Hills}, K. and {Hose}, K. and {Humbert}, T. and {Hurter}, F. and {Hynek}, T. and {Irzyk}, M. and {Jacobsen}, J. and {Jannetta}, A. L. and {Johnson}, K. and {Jóźwik-Wabik}, P. and {Kaeouach}, A. E. and {Kang}, W. and {Kiiskinen}, H. and {Kim}, T. and {Kivila}, Ü. and {Koch}, B. and {Kolb}, U. and {Kučáková}, H. and {Lai}, S.-P. and {Laloum}, D. and {Lasota}, S. and {Lewis}, L. A. and {Liakos}, G.-I. and {Libotte}, F. and {Lomoz}, F. and {Lopresti}, C. and {Majewski}, R. and {Malcher}, A. and {Mallonn}, M. and {Mannucci}, M. and {Marchini}, A. and {Mari}, J.-M. and {Marino}, A. and {Marino}, G. and {Mario}, J.-C. and {Marquette}, J.-B. and {Martínez-Bravo}, F. A. and {Mašek}, M. and {Matassa}, P. and {Michel}, P. and {Michelet}, J. and {Miller}, M. and {Miny}, E. and {Molina}, D. and {Mollier}, T. and {Monteleone}, B. and {Montigiani}, N. and {Morales-Aimar}, M. and {Mortari}, F. and {Morvan}, M. and {Mugnai}, L. V. and {Murawski}, G. and {Naponiello}, L. and {Naudin}, J.-L. and {Naves}, R. and {Néel}, D. and {Neito}, R. and {Neveu}, S. and {Noschese}, A. and {Öğmen}, Y. and {Ohshima}, O. and {Orbanic}, Z. and {Pace}, E. P. and {Pantacchini}, C. and {Paschalis}, N. I. and {Pereira}, C. and {Peretto}, I. and {Perroud}, V. and {Phillips}, M. and {Pintr}, P. and {Pioppa}, J.-B. and {Plazas}, J. and {Poelarends}, A. J. and {Popowicz}, A. and {Purcell}, J. and {Quinn}, N. and {Raetz}, M. and {Rees}, D. and {Regembal}, F. and {Rocchetto}, M. and {Rocci}, P.-F. and {Rockenbauer}, M. and {Roth}, R. and {Rousselot}, L. and {Rubia}, X. and {Ruocco}, N. and {Russo}, E. and {Salisbury}, M. and {Salvaggio}, F. and {Santos}, A. and {Savage}, J. and {Scaggiante}, F. and {Sedita}, D. and {Shadick}, S. and {Silva}, A. F. and {Sioulas}, N. and {Školník}, V. and {Smith}, M. and {Smolka}, M. and {Solmaz}, A. and {Stanbury}, N. and {Stouraitis}, D. and {Tan}, T.-G. and {Theusner}, M. and {Thurston}, G. and {Tifner}, F. P. and {Tomacelli}, A. and {Tomatis}, A. and {Trnka}, J. and {Tylšar}, M. and {Valeau}, P. and {Vignes}, J.-P. and {Villa}, A. and {Sureda}, A. Vives and {Vora}, K. and {Vrašt’ák}, M. and {Walliang}, D. and {Wenzel}, B. and {Wright}, D. E. and {Zambelli}, R. and {Zhang}, M. and {Zíbar}, M.},
  title = {ExoClock Project. III. 450 New Exoplanet Ephemerides from Ground and Space Observations},
  journal = {\apjs},
  year = {2023},
  volume = {265},
  pages = {4},
  doi = {10.3847/1538-4365/ac9da4},
}

\begin{appendix}
\onecolumn
\section{Sample parameters and timing data}
\label{app:tables}
\begin{table}[!htbp]
\caption{Sample and basic parameters of the \textit{Kepler} hot Jupiters.}
\label{tab:sample}
\centering
\renewcommand{\arraystretch}{1.15}
\begin{tabular}{lcccccc}
\hline\hline
Planet &
$M_{\rm p}$ ($M_{\rm Jup}$) &
$R_{\rm p}$ ($R_{\rm Jup}$) &
$M_\star$ ($M_\odot$) &
$R_\star$ ($R_\odot$) &
$a/R_\star$ &
Ref. \\
\hline

Kepler-5 b &
$2.111\pm0.086$ &
$1.426\pm0.051$ &
$1.374\pm0.059$ &
$1.793\pm0.062$ &
$6.450\pm0.025$ &
1 \\

Kepler-6 b &
$0.668\pm0.038$ &
$1.304\pm0.050$ &
$1.209\pm0.040$ &
$1.391\pm0.034$ &
$7.503\pm0.022$ &
1 \\

Kepler-7 b &
$0.441\pm0.043$ &
$1.622\pm0.013$ &
$1.359\pm0.064$ &
$1.966\pm0.013$ &
$6.637\pm0.021$ &
1 \\

Kepler-8 b &
$0.59\pm0.13$ &
$1.416\pm0.081$ &
$1.213\pm0.078$ &
$1.486\pm0.071$ &
$6.854\pm0.018$ &
1 \\

Kepler-12 b &
$0.432\pm0.041$ &
$1.754\pm0.077$ &
$1.166\pm0.055$ &
$1.483\pm0.062$ &
$8.019\pm0.027$ &
1 \\

Kepler-13 Ab &
$9.28\pm0.16$ &
$1.512\pm0.035$ &
$1.72\pm0.10$ &
$1.740\pm0.040$ &
$4.501\pm0.0040$ &
1 \\

Kepler-14 Ab &
$8.40\pm0.35$ &
$1.136\pm0.073$ &
$1.512\pm0.043$ &
$2.05\pm0.11$ &
$8.21\pm0.58$ &
2 \\

Kepler-15 b &
$0.660\pm0.090$ &
$0.960\pm0.070$ &
$1.018\pm0.052$ &
$0.992\pm0.070$ &
$12.8\pm1.5$ &
3 \\

Kepler-17 b &
$2.45\pm0.11$ &
$1.310\pm0.020$ &
$1.160\pm0.060$ &
$1.050\pm0.030$ &
$5.480\pm0.020$ &
4,5 \\

Kepler-40 b &
$2.20\pm0.40$ &
$1.170\pm0.040$ &
$1.480\pm0.060$ &
$2.130\pm0.060$ &
$8.110\pm0.060$ &
6 \\

Kepler-41 b &
$0.560\pm0.099$ &
$1.040\pm0.035$ &
$1.120\pm0.070$ &
$1.020\pm0.030$ &
$5.053\pm0.021$ &
1 \\

Kepler-43 b &
$3.23\pm0.26$ &
$1.219\pm0.065$ &
$1.320\pm0.090$ &
$1.420\pm0.070$ &
$6.975\pm0.047$ &
1,7 \\

Kepler-44 b &
$0.98\pm0.10$ &
$1.090\pm0.070$ &
$1.120\pm0.080$ &
$1.350\pm0.080$ &
$7.07\pm0.37$ &
5,8 \\

Kepler-45 b &
$0.505\pm0.090$ &
$0.96\pm0.11$ &
$0.590\pm0.060$ &
$0.55\pm0.11$ &
$10.6\pm1.0$ &
9 \\

Kepler-63 b &
$<0.378$ &
$0.545\pm0.018$ &
$0.984\pm0.040$ &
$0.901\pm0.027$ &
$19.120\pm0.080$ &
10 \\

Kepler-74 b &
$0.63\pm0.12$ &
$0.960\pm0.020$ &
$1.180\pm0.040$ &
$1.120\pm0.040$ &
$15.47\pm0.18$ &
8 \\

Kepler-75 b &
$10.10\pm0.40$ &
$1.050\pm0.030$ &
$0.910\pm0.040$ &
$0.890\pm0.020$ &
$19.77\pm0.45$ &
8 \\

Kepler-76 b &
$2.01\pm0.37$ &
$1.36\pm0.12$ &
$1.20\pm0.20$ &
$1.320\pm0.080$ &
$4.464\pm0.049$ &
1 \\

Kepler-77 b &
$0.430\pm0.032$ &
$0.960\pm0.016$ &
$0.950\pm0.040$ &
$0.990\pm0.020$ &
$9.764\pm0.055$ &
11 \\

Kepler-412 b &
$0.94\pm0.13$ &
$1.341\pm0.046$ &
$1.167\pm0.091$ &
$1.287\pm0.035$ &
$4.841\pm0.024$ &
1 \\

Kepler-422 b &
$0.43\pm0.13$ &
$1.15\pm0.11$ &
$1.150\pm0.060$ &
$1.24\pm0.12$ &
$14.2\pm1.9$ &
12,13 \\

Kepler-423 b &
$0.595\pm0.081$ &
$1.192\pm0.052$ &
$0.850\pm0.040$ &
$0.950\pm0.040$ &
$8.11\pm0.26$ &
14 \\

Kepler-424 b &
$1.03\pm0.13$ &
$0.890\pm0.080$ &
$1.010\pm0.054$ &
$0.940\pm0.056$ &
$10.060\ {\rm (derived)}$ &
12 \\

Kepler-426 b &
$0.340\pm0.080$ &
$1.090\pm0.030$ &
$0.910\pm0.060$ &
$0.920\pm0.020$ &
$9.670\pm0.060$ &
15 \\

Kepler-428 b &
$1.27\pm0.19$ &
$1.080\pm0.030$ &
$0.870\pm0.050$ &
$0.800\pm0.020$ &
$11.65\pm0.10$ &
15 \\

Kepler-433 b &
$2.82\pm0.52$ &
$1.45\pm0.16$ &
$1.46\pm0.17$ &
$2.26\pm0.25$ &
$6.44\pm0.62$ &
16 \\

Kepler-435 b &
$0.84\pm0.15$ &
$1.99\pm0.18$ &
$1.538\pm0.088$ &
$3.21\pm0.30$ &
$6.35\pm0.51$ &
16 \\

Kepler-447 b &
$1.37\pm0.48$ &
$1.65\pm0.59$ &
$1.00\pm0.21$ &
$1.05\pm0.19$ &
$20.41\pm0.36$ &
17 \\

\hline
\end{tabular}

\tablefoot{
References:
(1) \citet{Esteves2015};
(2) \citet{Buchhave2011};
(3) \citet{Endl2011};
(4) \citet{Desert2011};
(5) \citet{Bonomo2017};
(6) \citet{Santerne2011};
(7) \citet{Bonomo2012};
(8) \citet{Bonomo2015};
(9) \citet{Johnson2012};
(10) \citet{Sanchis2013};
(11) \citet{Gandolfi2013};
(12) \citet{Endl2014};
(13) \citet{Kokori2023};
(14) \citet{Gandolfi2015};
(15) \citet{Hebrard2014};
(16) \citet{Almenara2015};
(17) \citet{Lillo2015}.
For Kepler-424 b, $a/R_{\star}$ is derived from the published semimajor axis and stellar radius rather than directly reported in \citet{Endl2014}.
}
\end{table}

\begin{table}[!htbp]
\caption{Input transit parameters and usable transits in \textit{Kepler} LC, SC, and TESS data.}
\label{tab:ephemerides}
\centering
\renewcommand{\arraystretch}{1.15}
\begin{tabular}{lcccccc}
\hline\hline
Planet &
Reference $T_0$ (BKJD) &
$P$ (d) &
$\Delta_{14}$ (d) &
\textit{Kepler} LC &
\textit{Kepler} SC &
TESS \\
\hline
Kepler-5 b & 122.90144 & 3.5484657 & 0.190602 & 374 & 266 & 50 \\
Kepler-6 b & 121.486528 & 3.2346996 & 0.164201 & 334 & 218 & 48 \\
Kepler-7 b & 134.27687 & 4.8854892 & 0.216366 & 274 & 100 & 39 \\
Kepler-8 b & 121.11931 & 3.5224991 & 0.133727 & 385 & 260 & 46 \\
Kepler-12 b & 171.00915 & 4.4379629 & 0.195730 & 236 & 163 & 49 \\
Kepler-13 Ab & 120.56596 & 1.7635880 & 0.132000 & 758 & 584 & 158 \\
Kepler-14 Ab & 124.507037 & 6.79012033 & 0.259100 & 197 & 138 & 41 \\
Kepler-15 b & 136.328651 & 4.9427820 & 0.120624 & 243 & 76 & 54 \\
Kepler-17 b & 132.793576 & 1.48571092 & 0.094833 & 436 & 582 & 147 \\
Kepler-40 b & 138.151445 & 6.87316326 & 0.285833 & 148 & 0 & 0 \\
Kepler-41 b & 137.18104 & 1.8555577 & 0.098142 & 712 & 159 & 27 \\
Kepler-43 b & 132.41624 & 3.0240949 & 0.121917 & 440 & 290 & 68 \\
Kepler-44 b & 133.378969 & 3.24673248 & 0.134083 & 409 & 91 & 0 \\
Kepler-45 b & 131.537646 & 2.45524065 & 0.045463 & 532 & 343 & 64 \\
Kepler-63 b & 121.237981 & 9.43415263 & 0.120958 & 141 & 131 & 27 \\
Kepler-74 b & 134.346191 & 7.34071497 & 0.128400 & 179 & 38 & 11 \\
Kepler-75 b & 134.452105 & 8.88492372 & 0.086610 & 150 & 52 & 0 \\
Kepler-76 b & 133.54841 & 1.5449298 & 0.059970 & 683 & 0 & 19 \\
Kepler-77 b & 134.03058 & 3.57878055 & 0.121120 & 360 & 103 & 43 \\
Kepler-412 b & 133.02122 & 1.72086138 & 0.086542 & 370 & 171 & 105 \\
Kepler-422 b & 122.009187 & 7.89144809 & 0.178558 & 167 & 130 & 31 \\
Kepler-423 b & 133.354655 & 2.68432852 & 0.112243 & 459 & 189 & 53 \\
Kepler-424 b & 131.742972 & 3.31186426 & 0.058100 & 319 & 173 & 16 \\
Kepler-426 b & 133.631508 & 3.2175186 & 0.088000 & 404 & 120 & 41 \\
Kepler-428 b & 134.791658 & 3.52563245 & 0.108300 & 379 & 0 & 43 \\
Kepler-433 b & 131.98152 & 5.33408384 & 0.257400 & 250 & 57 & 8 \\
Kepler-435 b & 134.64037 & 8.60015312 & 0.375100 & 151 & 23 & 15 \\
Kepler-447 b & 121.671451 & 7.79430249 & 0.047290 & 141 & 0 & 25 \\
\hline
\end{tabular}
\tablefoot{Input transit parameters and the number of usable transits in \textit{Kepler} long-cadence (LC), short-cadence (SC), and TESS data for each system. BKJD denotes Barycentric \textit{Kepler} Julian Date, defined as BJD$-$2454833, where BJD is Barycentric Julian Date.}
\end{table}

\begin{table}[!htbp]
\caption{Mid-transit measurements for the 28 hot Jupiters.}
\label{tab:midtc}
\centering
\renewcommand{\arraystretch}{1.15}
\begin{tabular}{lccccc}
\hline\hline
Planet &
$T_c$ ($\mathrm{BJD}_{\mathrm{TDB}}$) &
$\sigma_{T_c}$ (d) &
Epoch &
Quarter &
Type \\
\hline
Kepler-5~b & 2454955.900605 & 0.000511 & 0 & 0 & 0 \\
Kepler-5~b & 2454959.449626 & 0.000997 & 1 & 0 & 0 \\
Kepler-5~b & 2454962.996928 & 0.001104 & 2 & 0 & 0 \\
Kepler-5~b & 2454966.546299 & 0.000290 & 3 & 1 & 0 \\
Kepler-5~b & 2454970.094245 & 0.001006 & 4 & 1 & 0 \\
\hline
\end{tabular}
\tablefoot{
This table is available in its entirety in machine-readable form.
A portion is shown here for guidance regarding its form and content.
Type = 0: \textit{Kepler} long cadence;
Type = 1: \textit{Kepler} short cadence;
Type = 2: TESS.
}
\end{table}

\begin{table}[!htbp]
\caption{Transit fit parameters and simulated ET data.}
\label{tab:transitparams}
\centering
\renewcommand{\arraystretch}{1.15}
\begin{tabular}{lccccccc}
\hline\hline
Planet &
$R_{\mathrm p}/R_\star$ &
$\rho_\star$ (g,cm$^{-3}$) &
$b$ &
$q_1$ &
$q_2$ &
$K_{\rm p}$ (mag) &
$N_{\mathrm{ET,sim}}$ \\
\hline

Kepler-5 b & $0.07868 \pm 0.00017$ & $0.403 \pm 0.010$ & $0.097 \pm 0.067$ & $0.25 \pm 0.02$ & $0.36 \pm 0.04$ & 13.369 & 412 \\

Kepler-6 b & $0.09310 \pm 0.00018$ & $0.782 \pm 0.013$ & $0.076 \pm 0.057$ & $0.39 \pm 0.02$ & $0.38 \pm 0.02$ & 13.303 & 451 \\

Kepler-7 b & $0.08225 \pm 0.00028$ & $0.237 \pm 0.006$ & $0.553 \pm 0.012$ & $0.32 \pm 0.03$ & $0.32 \pm 0.06$ & 12.885 & 299 \\

Kepler-8 b & $0.09340 \pm 0.00083$ & $0.492 \pm 0.015$ & $0.712 \pm 0.010$ & $0.37 \pm 0.07$ & $0.15 \pm 0.17$ & 13.563 & 414 \\

Kepler-12 b & $0.11790 \pm 0.00027$ & $0.486 \pm 0.007$ & $0.197 \pm 0.031$ & $0.33 \pm 0.02$ & $0.36 \pm 0.03$ & 13.438 & 329 \\

Kepler-13 Ab & $0.06473 \pm 0.00005$ & $0.369 \pm 0.003$ & $0.556 \pm 0.003$ & $0.28 \pm 0.01$ & $0.27 \pm 0.01$ & 9.958 & 828 \\

Kepler-14 Ab & $0.04547 \pm 0.00034$ & $0.166 \pm 0.016$ & $0.589 \pm 0.038$ & $0.35 \pm 0.06$ & $0.20 \pm 0.10$ & 12.128 & 215 \\

Kepler-15 b & $0.10166 \pm 0.00074$ & $0.805 \pm 0.020$ & $0.666 \pm 0.010$ & $0.37 \pm 0.08$ & $0.43 \pm 0.15$ & 13.758 & 296 \\

Kepler-17 b & $0.13059 \pm 0.00044$ & $1.572 \pm 0.032$ & $0.123 \pm 0.067$ & $0.32 \pm 0.03$ & $0.39 \pm 0.04$ & 14.141 & 984 \\

Kepler-40 b & $0.05782 \pm 0.00132$ & $0.126 \pm 0.030$ & $0.538 \pm 0.128$ & $0.47 \pm 0.23$ & $0.18 \pm 0.21$ & 14.588 & 213 \\

Kepler-41 b & $0.09989 \pm 0.00086$ & $0.737 \pm 0.021$ & $0.677 \pm 0.011$ & $0.43 \pm 0.11$ & $0.25 \pm 0.15$ & 14.465 & 787 \\

Kepler-43 b & $0.08495 \pm 0.00059$ & $0.724 \pm 0.027$ & $0.640 \pm 0.014$ & $0.39 \pm 0.08$ & $0.22 \pm 0.12$ & 13.958 & 483 \\

Kepler-44 b & $0.07816 \pm 0.00110$ & $0.722 \pm 0.079$ & $0.559 \pm 0.053$ & $0.62 \pm 0.19$ & $0.15 \pm 0.13$ & 14.678 & 450 \\

Kepler-45 b & $0.17890 \pm 0.00239$ & $4.065 \pm 0.102$ & $0.538 \pm 0.021$ & $0.56 \pm 0.23$ & $0.28 \pm 0.16$ & 15.979 & 595 \\

Kepler-63 b & $0.06375 \pm 0.00097$ & $1.398 \pm 0.104$ & $0.748 \pm 0.017$ & $0.25 \pm 0.07$ & $0.52 \pm 0.32$ & 11.582 & 155 \\

Kepler-74 b & $0.08904 \pm 0.00130$ & $1.257 \pm 0.097$ & $0.696 \pm 0.025$ & $0.36 \pm 0.13$ & $0.19 \pm 0.28$ & 14.412 & 199 \\

Kepler-75 b & $0.11496 \pm 0.00298$ & $10.466 \pm 1.550$ & $0.267 \pm 0.171$ & $0.47 \pm 0.30$ & $0.34 \pm 0.28$ & 15.264 & 165 \\

Kepler-76 b & $0.11471 \pm 0.00678$ & $0.733 \pm 2.357$ & $0.895 \pm 0.387$ & $0.97 \pm 0.05$ & $0.98 \pm 0.06$ & 13.308 & 946 \\

Kepler-77 b & $0.09753 \pm 0.00050$ & $1.346 \pm 0.045$ & $0.353 \pm 0.033$ & $0.45 \pm 0.06$ & $0.34 \pm 0.05$ & 13.938 & 409 \\

Kepler-412 b & $0.10261 \pm 0.00144$ & $0.720 \pm 0.018$ & $0.800 \pm 0.006$ & $0.36 \pm 0.07$ & $0.22 \pm 0.28$ & 14.309 & 849 \\

Kepler-422 b & $0.09575 \pm 0.00047$ & $0.759 \pm 0.024$ & $0.500 \pm 0.019$ & $0.31 \pm 0.05$ & $0.43 \pm 0.09$ & 13.435 & 185 \\

Kepler-423 b & $0.12380 \pm 0.00049$ & $1.381 \pm 0.028$ & $0.351 \pm 0.022$ & $0.37 \pm 0.05$ & $0.40 \pm 0.05$ & 14.290 & 544 \\

Kepler-424 b & $0.09302 \pm 0.00386$ & $1.779 \pm 0.140$ & $0.926 \pm 0.012$ & $0.47 \pm 0.10$ & $0.41 \pm 0.10$ & 14.256 & 441 \\

Kepler-426 b & $0.11758 \pm 0.00132$ & $1.615 \pm 0.060$ & $0.724 \pm 0.012$ & $0.43 \pm 0.12$ & $0.16 \pm 0.23$ & 14.835 & 454 \\

Kepler-428 b & $0.14788 \pm 0.00166$ & $0.963 \pm 0.023$ & $0.700 \pm 0.009$ & $0.29 \pm 0.11$ & $0.77 \pm 0.26$ & 15.224 & 414 \\

Kepler-433 b & $0.06314 \pm 0.00049$ & $0.214 \pm 0.020$ & $0.219 \pm 0.128$ & $0.34 \pm 0.08$ & $0.26 \pm 0.09$ & 14.463 & 274 \\

Kepler-435 b & $0.06192 \pm 0.00033$ & $0.105 \pm 0.006$ & $0.322 \pm 0.064$ & $0.41 \pm 0.07$ & $0.17 \pm 0.06$ & 13.643 & 170 \\

Kepler-447 b & $0.16598 \pm 0.02914$ & $2.530 \pm 0.219$ & $1.082 \pm 0.034$ & $0.53 \pm 0.10$ & $0.40 \pm 0.09$ & 12.394 & 188 \\
\hline

\end{tabular}

\tablefoot{$K_{\rm p}$ values were obtained from the NASA Exoplanet Archive and denote the \textit{Kepler}-band magnitude, not the near-infrared $K$-band magnitude. $N_{\mathrm{ET,sim}}$ is the number of simulated ET transits generated using these parameters.}

\end{table}

\begin{table}[!htbp]
\caption{Timing analysis results for the combined \textit{Kepler}, TESS, and simulated ET dataset.}
\label{tab:timing}
\centering
\renewcommand{\arraystretch}{1.15}
\begin{tabular}{lcccccccc}
\hline\hline
Planet &
\makecell{$T_0$\\(BJD$-$2454833)} &
\makecell{$\sigma_{T_0}$\\($10^{-5}$ d)} &
\makecell{$P$\\(d)} &
\makecell{$\sigma_P$\\($10^{-8}$ d)} &
\makecell{$\dot{P}_{\rm theory}$\\(ms\,yr$^{-1}$)} &
\makecell{$\dot{P}_{\rm KTE}$\\(ms\,yr$^{-1}$)} &
\makecell{$\sigma_{\dot{P},{\rm KTE}}$\\(ms\,yr$^{-1}$)} &
$\Delta\mathrm{BIC}_{\rm KTE}$ \\
\hline
Kepler-5 b & 122.90141 & 3.0 & 3.54846564 & 15 & -0.35 & 0.73 & 0.90 & -5.9 \\
Kepler-6 b & 121.48659 & 2.0 & 3.23469927 & 10 & -0.06 & -0.23 & 0.56 & -6.5 \\
Kepler-7 b & 124.50594 & 3.4 & 4.88548872 & 23 & -0.06 & 0.1 & 1.4 & -6.4 \\
Kepler-8 b & 121.11946 & 3.5 & 3.52249827 & 16 & -0.08 & 0.95 & 0.91 & -5.7 \\
Kepler-12 b & 122.19156 & 2.0 & 4.43796288 & 12 & -0.03 & 0.48 & 0.41 & -5.1 \\
Kepler-13 Ab & 122.32962 & 1.0 & 1.76358761 & 1.0 & -7.44 & -7.24 & 0.08 & 8165.2 \\
Kepler-14 Ab & 124.50789 & 8.0 & 6.79012202 & 73 & -0.38 & 4.3 & 3.0 & -4.1 \\
Kepler-15 b & 136.32941 & 4.0 & 4.94278360 & 31 & 0.00 & -1.3 & 1.6 & -5.7 \\
Kepler-17 b & 132.79360 & 1.0 & 1.48571088 & 2.0 & -1.09 & -1.01 & 0.13 & 54.9 \\
Kepler-40 b & 138.15319 & 24 & 6.87316530 & 210 & -0.11 & 20 & 12 & -3.3 \\
Kepler-41 b & 131.61435 & 3.0 & 1.85555760 & 9.0 & -0.39 & -1.18 & 0.47 & -1.2 \\
Kepler-43 b & 132.41692 & 4.0 & 3.02409222 & 13 & -0.38 & -0.01 & 0.87 & -6.7 \\
Kepler-44 b & 133.37920 & 9.0 & 3.24673169 & 38 & -0.13 & -3.0 & 1.9 & -4.7 \\
Kepler-45 b & 131.53783 & 3.0 & 2.45524077 & 8.1 & -0.02 & 0.23 & 0.51 & -6.9 \\
Kepler-63 b & 121.23835 & 10 & 9.43415268 & 69 & 0.00 & 6.1 & 3.5 & -3.0 \\
Kepler-74 b & 134.34590 & 8.0 & 7.34071276 & 71 & 0.00 & -3.3 & 4.5 & -5.5 \\
Kepler-75 b & 134.45205 & 13 & 8.88492100 & 150 & -0.01 & -0.1 & 5.7 & -5.7 \\
Kepler-76 b & 132.00343 & 3.0 & 1.54492981 & 1.0 & -2.41 & -2.05 & 0.49 & 11.4 \\
Kepler-77 b & 134.03036 & 3.0 & 3.57878087 & 18 & -0.01 & 0.11 & 0.87 & -6.7 \\
Kepler-412 b & 133.02099 & 3.0 & 1.72086127 & 9.0 & -0.77 & -0.38 & 0.43 & -6.5 \\
Kepler-422 b & 122.00995 & 4.0 & 7.89144727 & 39 & 0.00 & 0.4 & 2.5 & -6.0 \\
Kepler-423 b & 136.03919 & 2.0 & 2.68432829 & 8.0 & -0.05 & 0.09 & 0.43 & -6.9 \\
Kepler-424 b & 131.74274 & 8.0 & 3.31186442 & 37 & -0.03 & 1.3 & 1.9 & -6.2 \\
Kepler-426 b & 133.63197 & 4.0 & 3.21751876 & 16 & -0.01 & 1.52 & 0.98 & -4.5 \\
Kepler-428 b & 134.79169 & 4.0 & 3.52563256 & 16 & -0.02 & -0.7 & 1.0 & -6.2 \\
Kepler-433 b & 131.98147 & 13 & 5.33408510 & 87 & -0.44 & -0.6 & 5.1 & -6.3 \\
Kepler-435 b & 134.64148 & 13 & 8.60015490 & 140 & -0.13 & 5.3 & 8.9 & -5.4 \\
Kepler-447 b & 121.67178 & 19 & 7.79430179 & 97 & 0.00 & 2.5 & 6.1 & -5.6 \\
\hline
\end{tabular}
\tablefoot{
Results of the timing analysis combining the observed \textit{Kepler} and \textit{TESS} data with four years of simulated ET observations.
$T_0$ and $P$ are the fitted reference transit epoch and orbital period.
$\dot{P}_{\rm theory}$ is the input orbital period derivative adopted in the ET simulations, calculated assuming $\log Q'_\star = 5.7$.
$\dot{P}_{\rm KTE}$ and $\sigma_{\dot{P},{\rm KTE}}$ are the recovered orbital period derivative and its $1\sigma$ uncertainty from fitting the combined dataset.
$\Delta\mathrm{BIC}_{\rm KTE}$ denotes $\mathrm{BIC}_{\rm lin}-\mathrm{BIC}_{\rm quad}$.}
\end{table}

\end{appendix}
\end{document}